\documentclass[twocolumn]{aastex62}
\usepackage{epsfig,amsmath,amsfonts,amssymb,graphicx,subfigure,enumitem,url,natbib}
\received{.........}
\revised{.........}
\accepted{........}
\submitjournal{AAS}
\shorttitle{Variable stars in NGC 4147}
\shortauthors{Sneh Lata et al.}
 
\begin{document}
\title{VR CCD photometry of variable stars in globular cluster NGC 4147}

\email{sneh@aries.res.in}

\author{Sneh Lata} 
\affil{Aryabhatta Research Institute of Observational Sciences, Manora Peak, Nainital 263002, Uttarakhand, India }
\author{A. K. Pandey}
\affil{Aryabhatta Research Institute of Observational Sciences, Manora Peak, Nainital 263002, Uttarakhand, India }
\author{J. C. Pandey}
\affil{Aryabhatta Research Institute of Observational Sciences, Manora Peak, Nainital 263002, Uttarakhand, India }
\author{R. K. S. Yadav}
\affil{Aryabhatta Research Institute of Observational Sciences, Manora Peak, Nainital 263002, Uttarakhand, India }
\author{Shashi B. Pandey}
\affil{Aryabhatta Research Institute of Observational Sciences, Manora Peak, Nainital 263002, Uttarakhand, India }
\author{Aashish Gupta}
\affil{PDPM Indian Institute of Information Technology Design \& Manufacturing Jabalpur Dumna Airport Road, Dumna - 482005 Jabalpur, Madhya Pradesh, India }
\author{Tarun Bangia}
\affil{Aryabhatta Research Institute of Observational Sciences, Manora Peak, Nainital 263002, Uttarakhand, India }
\author{Hum Chand}
\affil{Aryabhatta Research Institute of Observational Sciences, Manora Peak, Nainital 263002, Uttarakhand, India }
\author{Mukesh K. Jaiswar}
\affil{Aryabhatta Research Institute of Observational Sciences, Manora Peak, Nainital 263002, Uttarakhand, India }
\author{Yogesh C. Joshi}
\affil{Aryabhatta Research Institute of Observational Sciences, Manora Peak, Nainital 263002, Uttarakhand, India }
 \author{Mohit Joshi}
\affil{Aryabhatta Research Institute of Observational Sciences, Manora Peak, Nainital 263002, Uttarakhand, India }
 \author{Brijesh Kumar}
\affil{Aryabhatta Research Institute of Observational Sciences, Manora Peak, Nainital 263002, Uttarakhand, India }
 \author{T. S. Kumar}
\affil{Aryabhatta Research Institute of Observational Sciences, Manora Peak, Nainital 263002, Uttarakhand, India }
 \author{Biman J. Medhi}
\affil{Aryabhatta Research Institute of Observational Sciences, Manora Peak, Nainital 263002, Uttarakhand, India }
\affil{Department of Physics, Gauhati University, Jalukbari, Guwahati  781014}
 \author{Kuntal Misra}
\affil{Aryabhatta Research Institute of Observational Sciences, Manora Peak, Nainital 263002, Uttarakhand, India }
 \author{Nandish Nanjappa}
\affil{Aryabhatta Research Institute of Observational Sciences, Manora Peak, Nainital 263002, Uttarakhand, India }
 \author{Jaysreekar Pant}
\affil{Aryabhatta Research Institute of Observational Sciences, Manora Peak, Nainital 263002, Uttarakhand, India }
 \author{Purushottam}
\affil{Aryabhatta Research Institute of Observational Sciences, Manora Peak, Nainital 263002, Uttarakhand, India }
 \author{B. Krishna Reddy}
\affil{Aryabhatta Research Institute of Observational Sciences, Manora Peak, Nainital 263002, Uttarakhand, India }
 \author{Sanjeet Sahu}
\affil{Aryabhatta Research Institute of Observational Sciences, Manora Peak, Nainital 263002, Uttarakhand, India }
 \author{Saurabh Sharma}
\affil{Aryabhatta Research Institute of Observational Sciences, Manora Peak, Nainital 263002, Uttarakhand, India }
 \author{Wahab Uddin}
\affil{Aryabhatta Research Institute of Observational Sciences, Manora Peak, Nainital 263002, Uttarakhand, India }
 \author{Shobhit Yadav }
\affil{Aryabhatta Research Institute of Observational Sciences, Manora Peak, Nainital 263002, Uttarakhand, India }
 

\begin{abstract}
We present results of a search for variable stars in a region of the globular cluster NGC 4147 based on photometric observations with 4K$\times$4K CCD imager mounted at the axial port of the recently installed 3.6 m Devasthal optical telescope at Aryabhatta Research Institute of
Observational Sciences, Nainital, India.
We performed time series photometry of NGC 4147 in $V$ and $R$ bands, and identified 42 periodic 
variables in the region of NGC 4147, 28 of which have been detected for the first time.  
Seventeen variable stars are located within the half light radius $\lesssim$ 0.48 arcmin, of which 10 stars are newly identified variables.
Two of 10 variables are located within the core radius $\lesssim$ 0.09 arcmin.
Based on the location in the $V/(V-R)$ colour magnitude diagram and variability characteristics,
7, 8, 5 and 1 newly identified probable member variables are classified as RRc, EA/E, EW and SX Phe, respectively. 
The metallicity of NGC 4147 estimated from light curves of RRab and RRc stars with the help of Fourier decomposition is found to be
characteristics of Oosterhoff II.
The distance derived using light curves of RRab stars is consistent with that obtained from the observed $V/(V-R)$ colour-magnitude diagram. 
\end{abstract}

\section{Introduction}
The study of variable stars in globular clusters has been carried out by several groups using 
different methods and techniques (e.g., Clement et al. 2001, Pritzl et al. 2002, Layden et al. 2003, Clementini et al. 2005, Corwin et al. 2006, Baker et al. 2007 and Kopacki et al. 2008).  Globular clusters normally consist of low mass population II stars and provide an ideal environment for the search of RR Lyrae stars.  The early investigations of globular clusters show that the more than 90\% of the known variables were of the RR Lyrae type.  With the advent of new  technology, other types of variables in the globular clusters are also discovered. Even after new discoveries of variables,  RR Lyrae stars still dominate the variability populations in globular clusters. They now constitute less than 70\% of the known variables (see Clement 2017).

NGC 4147 is a relatively small globular cluster of radius 0.48 arcmin, which is located at 21 kpc from the Galactic center and 19 kpc from the Sun and has low metallicity $[Fe/H]$ = -1.83 (Harris 1996; 2010 edition). The 0.48 arcmin radius given in the Harris catalogue (Harris 1996; 2010 edition) refers to the half light radius of the cluster.  The core and tidal radii of the cluster are as 0.09 and 6.1 arcmin, respectively ((Harris 1996; 2010 edition). The reddening towards the cluster NGC 4147 is found to be $E(B-V)$=0.02 mag (Harris 1996; 2010 edition). Previous studies by Castellani \& Quarta (1987), Arellano Ferro et al. (2004) and Stetson et al. (2005) have shown that, although NGC 4147 is a relatively metal poor cluster with a predominantly blue horizontal branch (HB) in its colour-magnitude diagram (CMD), its RR Lyrae have short periods, characteristic of more metal rich Oosterhoff type I clusters. Meanwhile, the large proportion of RRc variables is typical of an Oosterhoff II cluster.

Up to now, the only known variables in NGC 4147 are RR Lyrae.
With much better plate scale and the large aperture telescope, it should be  possible to identify more RR Lyrae and other variable stars in the crowded central region of the globular cluster NGC 4147. Aiming this,
 the observations of globular cluster NGC 4147 were taken using the 4k$\times$4k CCD camera, the first light imaging instrument mounted at the axial port of recently installed 3.6-m Devasthal optical telescope (DOT) at Aryabhatta Research Institute of
Observational Sciences (ARIES), Devasthal campus, India (Pandey et al. 2017). 
A time series photometry was carried out on 6 nights from 23 March 2017 to 9 April 2017 in order to search the photometric variables in the cluster NGC 4147.
With these new  observations, we present precision of photometry and compare the present results with the previously published work. The paper is organised as follows. In the next section, we present observations and data reduction. In subsequent sections we present variable identification and discuss their association to the cluster on the basis of $V/(V-R)$ CMD, proper motions and geometric probabilities. We discuss period determination of the known and newly detected variable stars in subsequent section. Finally, we present study on RR Lyrae variables as well as other variables detected in the region of the globular cluster NGC 4147.

\begin{figure*}
\hbox{
\hspace{-1.8cm}
\includegraphics[width=12cm]{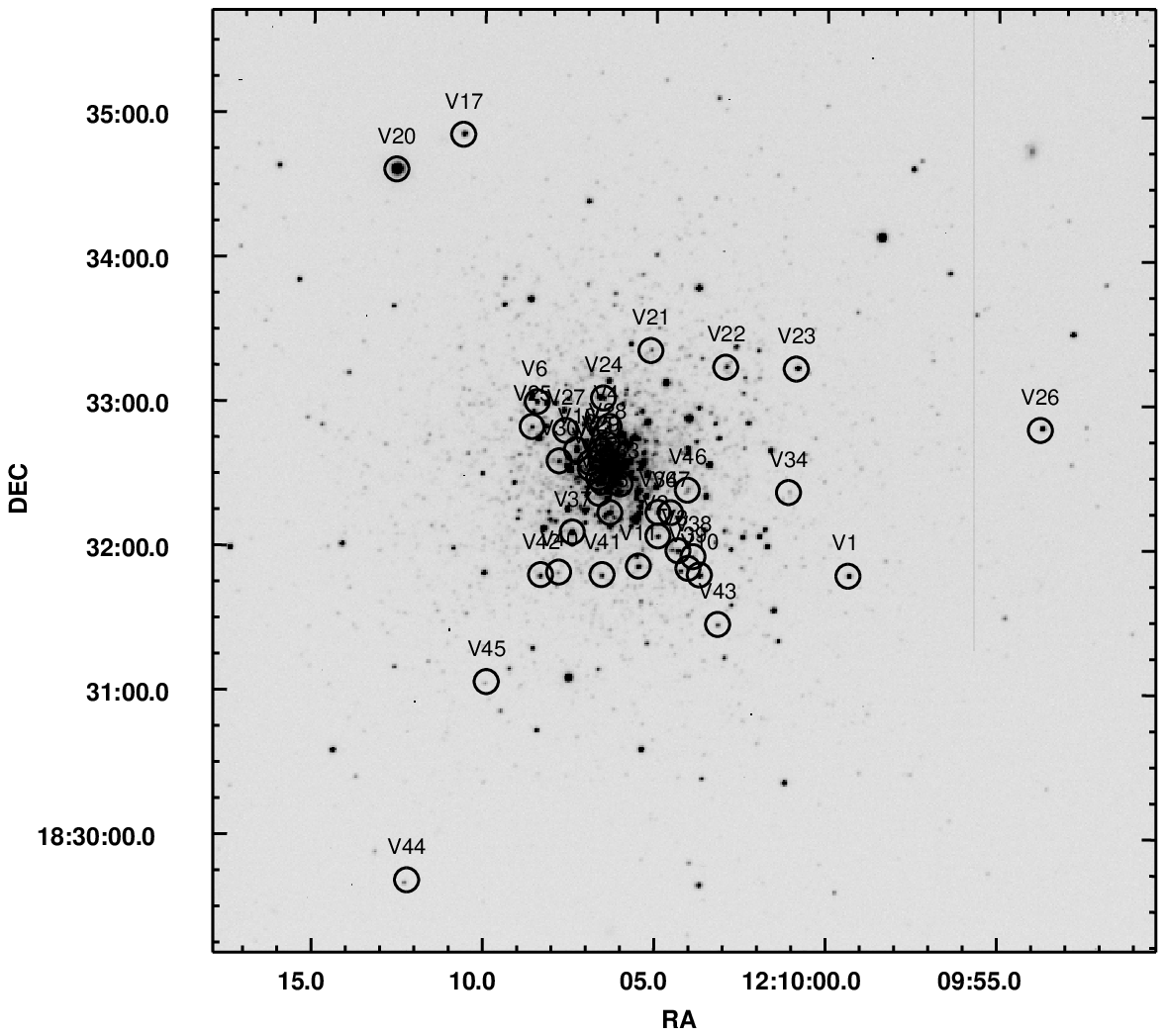}
\hspace{-2.8cm}
\includegraphics[width=10cm,height=7.5cm]{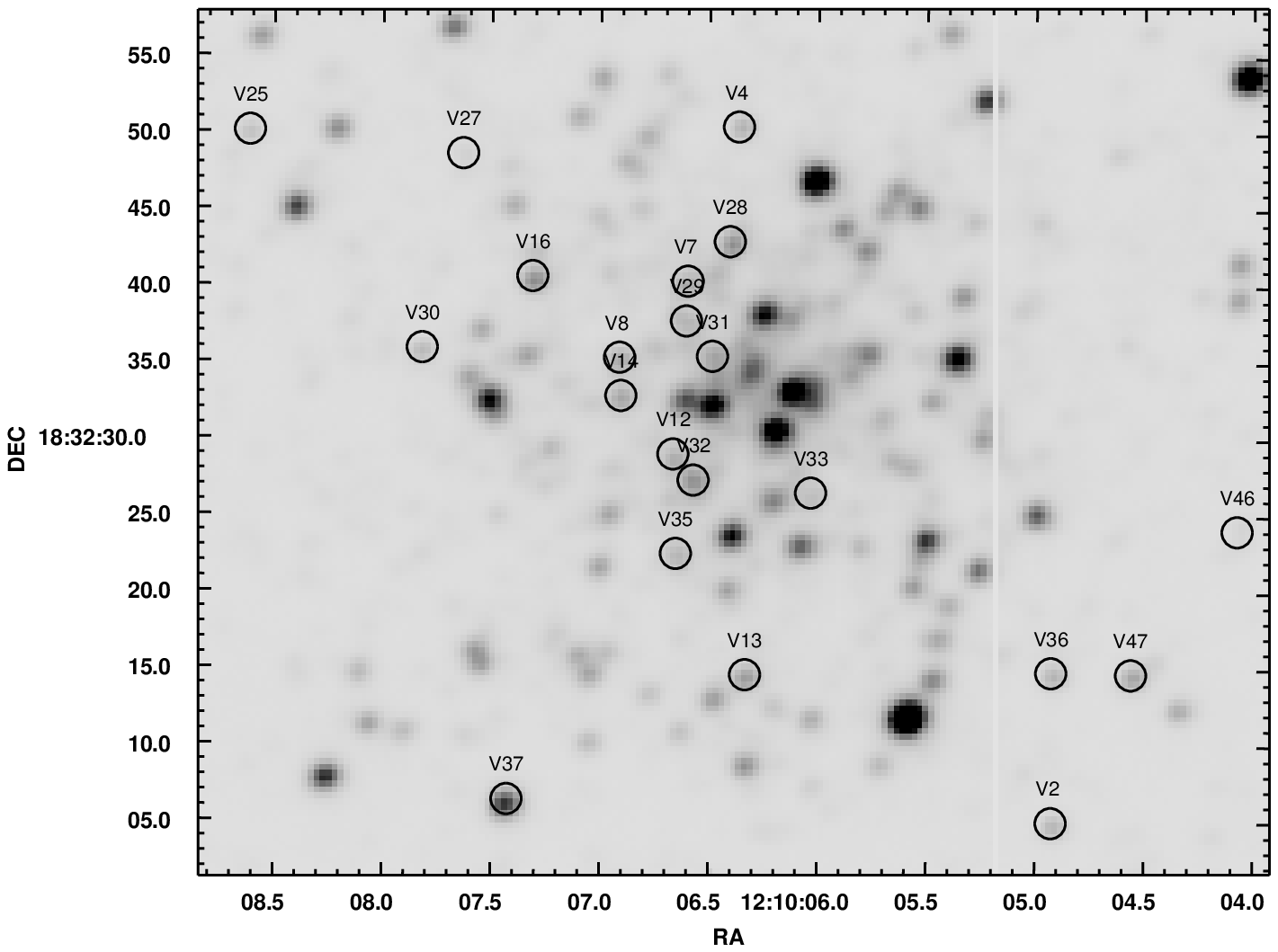}
}
\caption{ Left panel shows observed region of NGC 4147 in $V$-band while right panel represent central region of the cluster. 
The variable candidates detected in the present work are encircled and labeled with IDs.}
\end{figure*}

\begin{table}
\caption{Log of the observations. N and Exp. represent number of frames obtained and exposure time, respectively. \label{tab:obsLog}}
\begin{tabular}{lclcc}
\hline
S. No.&Date of        &Object&{\it V}                &{\it R}                               \\
         &observations&         &(N$\times$Exp.)&(N$\times$Exp.) \\
\hline
1 & 23 March 2017  & NGC 4147&- &5$\times$300s \\
2 & 24 March 2017  & NGC 4147&45$\times$40s &45$\times$40s\\
3 & 25 March 2017  & NGC 4147&39$\times$50s &39$\times$50s \\
4 & 28 March 2017  & NGC 4147&80$\times$50s &38$\times$50s \\
5 & 08 April 2017  & NGC 4147&58$\times$30s &58$\times$30s \\
6 & 09 April 2017  & NGC 4147&117$\times$30s&117$\times$30s\\
\hline
\end{tabular}
\end{table}

\section{Observations and Data Reduction}
The CCD observations were carried using 3.6 m DOT 
(f/9 Alt-azimuth mounting) at Aryabhatta Research Institute of
Observational Sciences (ARIES), Devasthal (Nainital), India. 
The telescope is equipped with 4k$\times$4k CCD imager which is mounted in the main port of the telescope.
The CCD imager has a pixel size of 15 $\micron$ 
and a plate scale of 0.095 arcsec per pixel, covering approximately 6.23$\times$6.23 arcmin$^2$
field of view over the sky.
The observations were taken in the 4$\times$4 binning mode with 1000 kHz readout.
We have made observations in $V$ and $R$ bands on  6  nights  between
2017 March 23  and April 9. In total, we have collected 339 and
302 frames in the $V$ and $R$
bands, respectively. 
The exposure times vary from 30 to 50 seconds.
The FWHM of the observations varies between $\sim$ 0.7 to 1.0 arcsec.
The sky brightness was found to be  high due to the bright Moon on two nights. 
Bias and twilight flats were also taken along with the target field.
The log of observations is given in Table 1.

The  pre-processing  of  the  images  was  performed  using IRAF\footnote{IRAF is distributed by the National Optical Astronomy Observatory, which is operated by the Association of Universities for Research in Astronomy (AURA) under cooperative agreement with the National Science Foundation}.
The instrumental magnitudes of the stars were obtained using the DAOPHOT package (Stetson 1987).
Both aperture and  PSF photometry were performed as PSF photometry 
gives better results for the crowded region.
For the PSF photometry, we have selected bright isolated stars across the 
field to construct a characteristic point-spread function (PSF) for the images. The PSF photometry of all the sources was obtained using the ALLSTAR task. 
 Fig. 1 shows the observed field of globular cluster NGC 4147.
The data of $V$ and $R$ bands were merged by matching the coordinates using a radial matching tolerance of 1 arcsec. 
We have detected 1057 stars in both $V$ and $R$ bands. DAOMATCH (Stetson 1992) routine of DAOPHOT  was used to find the translation, rotation and scaling solutions between different photometry files, whereas we have used DAOMASTER (Stetson 1992) to match the point sources. DAOMASTER was also used to remove the effects of frame-to-frame flux variation
due to airmass and exposure time. This task makes the mean flux level of each frame equal to the reference frame by an additive constant.
The first target image on 25 March 2017 was taken as the reference frame.

The intensity weighted mean instrumental magnitudes in $V$ and $R$ bands given by DAOMASTER (Stetson 1992) were transformed into the standard ones using the 
photometric data of standard stars marked in the field of NGC 4147 by Arellano Ferro et al. (2004). We have also used data of few stars taken as standard by Stetson et al. (2005) to calibrate present instrumental magnitudes. 
The following transformation equations were obtained
\begin{eqnarray}
V = v+(-0.198\pm0.021)\times (V-R) + 4.800\pm0.022   \nonumber\\
V-R =(1.036\pm0.032)\times (v-r)-0.0874\pm0.019   \nonumber
\end{eqnarray}

where $V$ and $R$ denote standard magnitudes, and $v$ and $r$
are the instrumental magnitudes in $V$ and $R$ bands, respectively. 
The X and Y coordinates of the reference image were transformed into Right ascension and Declination using CCMAP and CCTRAN tasks available in IRAF.
The average photometric error of the data was estimated using the observations of each star. The average photometric error of stars along with identified variables 
as a function of standard magnitude in $V$ and $R$ bands is shown in Fig. 2. 
We have plotted only those stars which are having photometric error $\lesssim$0.02 mag up to $V$$\sim$ 18 mag. Whereas all the identified variables, irrespective of the photometric errors, are plotted in Fig. 2.
The photometric errors in $(V-R)$ colours of the stars have been calculated as $\sigma{_{V-R}}=sqrt(\sigma{_V}^2+\sigma{_R}^2)$.

\begin{figure}
\includegraphics[width=8cm]{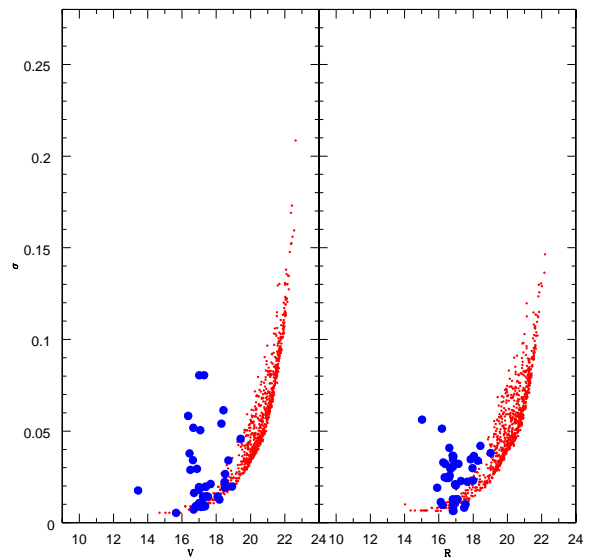}
\caption{Mean photometric errors of stars in the cluster region NGC 4147 given by DAOPHOT as a function of standard magnitude in $V$ and $R$ bands.
Stars having unreliable photometry up to $V$$\sim$18.0 mag have not been plotted. The larger filled circles display variable stars.   
}
\end{figure}

\subsection{Comparison with previous photometry}
We have compared present photometric results with the previous photometry given by Stetson et al. (2005) and Wang et al. (2000).
Data have been downloaded from the Vizier catalog for both previous photometries. The catalog by Stetson et al. (2005) lists 
$BV$ data for 91 stars while Wang et al. (2000) presented photometry for 115 stars. 
Cross-identification yields 79 and 94 common stars with 
Stetson et al. (2005) and Wang et al. (2000), respectively.
Fig. 3 shows the difference $\Delta$ (in the sense present minus previous photometry) in $V$ and $(V-R)$ between our photometry and previous photometries. 
The filled circles represent the difference between present $V$ magnitudes and those by Stetson et al. (2005) while open circles show the difference between present photometry and the photometry by Wang et al. (2000). Fig. 3 indicates that present $V$ magnitudes match well with those given by 
Stetson et al. (2005), whereas
comparison of present $V$ magnitudes with those obtained by Wang et al. (2000) indicates fair agreement up to $V$$\sim$ 16.0 mag. The present $V$ magnitudes
become fainter than that of Wang et al. (2000) after $V$$\sim$ 16.0 mag. The $(V-R)$ colours obtained by Wang et al. (2000) are redder and become even more redder with increase of $V$ magnitudes.

\subsection{Variables identification}
The corrected instrumental magnitudes of variables provided by the DAOMASTER  were converted into 
standard ones by obtaining transformation equations using standard stars. 
We have obtained transformation equations as follows
\begin{eqnarray}
V = (1.000\pm0.002)\times v+4.845\pm0.010   \nonumber\\
R = (0.998\pm0.007)\times r+4.797\pm0.005   \nonumber
\end{eqnarray}

The colour term was not used in the above equations as it is found insignificant (see also Arellano Ferro et al. 2004). 
We have identified variable stars by inspecting  their light curves visually. 
The light curves of all the stars were displayed using a code written in IDL and each of them
 was visually inspected, and those showed periodic/regular brightness variation were selected.
We have 
detected a total of 42 variables in the region of NGC 4147.
Among these 42 variables, 28 variables have been detected for the first time and 
these variables are termed as
newly identified variables.
Stars V20 and V31 could not be detected in the $R$ band while we could not detect V12 in the $V$ band. 
The optical data of variable stars along with their identification numbers are listed in Table 2. 
Column 1 of Table 2 gives the identification number of a variable.
We have adopted the same nomenclature for the known variable stars which have been used by earlier studies (see Stetson et al. 2005).
The identification numbers for new variable stars are assigned in continuation of the past
work.
The positions of identified variables are shown by the circles in Fig. 1.
This clearly shows that most of the variables identified in the present study are found to be located towards the cluster center.
To understand the spatial distribution of identified variables in a better way, the radial distance ($r$) of each variable is
determined by estimating the cluster center (RA=12h:10m:6.31 and DEC=+18d:32m:33.42s) and given in Table 2.
Two newly identified variables V29 and V31 are located within the core radius $\lesssim$ 0.09 arcmin.
Seventeen variable stars including V29 and 31 are found to be distributed within the half light radius $\lesssim$ 0.48 arcmin, of which 10 stars namely V24, V27, V28, V29, V30, 31, V32, V33, V35 and V36 are newly identified variables.
 Data of all the detected variable stars in present work is available online. The format of data is as follows. The first, second and third
column represent JD, magnitude and error, respectively.

\begin{figure}
\includegraphics[width=8cm]{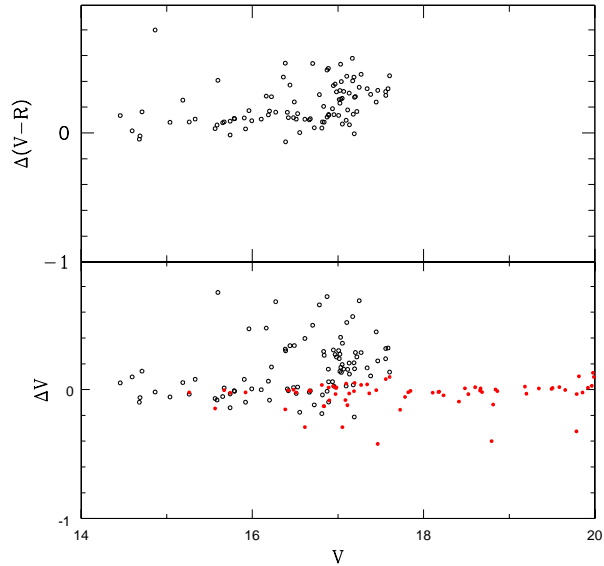}
\caption{Comparison of the present photometry and previous photometry given by Stetson et al. (2005) and Wang et al. (2000).
Filled circles represent difference between the present and Stetson et al. (2005) data and open circles shows difference between the present and  Wang et al. (2000) data. The $\Delta$ represents difference between the present and previous photometries.}
\end{figure}

\begin{table*}
\caption{The photometric data, period, amplitude, radial distance and classification of 42 variables in the region of NGC 4147.
}
\tiny
\begin{tabular}{llllccclll}
\hline
ID  &  RA(2000)&   Dec(2000)&      $V$&   $V-R$&     $r$ & $Amp_{V}$&  $Amp_{R}$ & Period& Class. \\
    &          &            &      (mag)&    (mag) & (arcmin) &   (mag)& (mag) &(days)&        \\
\hline
V1    &       182.4973880 &  18.5302222   &   17.350$\pm$0.0089   &      0.358$\pm$0.0132  & 1.814 &    1.149    & 0.919   & 0.500390 &RRab  \\
V2    &       182.5205278 &  18.5347222   &   17.190$\pm$0.0089   &      0.337$\pm$0.0131  & 0.582 &    1.170    & 1.123   & 0.330507 &RRab  \\
V3    &       182.5180833 &  18.5330278   &   16.890$\pm$0.0090   &      0.044$\pm$0.0132  & 0.747 &    0.584    & 0.515   & 0.281159 &RRc  \\
V4    &       182.5265278 &  18.5473333   &   17.220$\pm$0.0143   &      0.402$\pm$0.0216  & 0.280 &    0.709    & 0.678   & 0.299347 &RRc  \\
V6    &       182.5352778 &  18.5501111   &   17.248$\pm$0.0090   &      0.406$\pm$0.0131  & 0.677 &    1.019    & 0.872   & 0.609663 &RRab \\
V7    &       182.5275000 &  18.5445278   &   16.659$\pm$0.0518   &      0.321$\pm$0.0755  & 0.128 &    1.090    & 1.029   & 0.514077 &RRab \\
V8    &       182.5288056 &  18.5431389   &   16.872$\pm$0.0294   &      0.283$\pm$0.0403  & 0.142 &    0.518    & 0.425   & 0.275948 &RRc  \\
V10   &       182.5154722 &  18.5302500   &   17.134$\pm$0.0090   &      0.331$\pm$0.0120  & 0.971 &    0.405    & 0.413   & 0.352831 &RRc  \\
V11   &       182.5228889 &  18.5312222   &   17.007$\pm$0.0107   &      0.322$\pm$0.0156  & 0.717 &    0.489    & 0.432   & 0.387690 &RRc  \\
V12   &       182.5277778 &  18.5413889   &   -                   &      -                 & 0.113 &   -         & 1.319   & 0.504250 &RRab \\
V13   &       182.5263889 &  18.5373889   &   16.704$\pm$0.0162   &      0.243$\pm$0.0240  & 0.318 &    0.499    & 0.442   & 0.406708 &RRc  \\
V14   &       182.5287778 &  18.5424444   &   16.634$\pm$0.0341   &      0.257$\pm$0.0508  & 0.139 &    0.338    & 0.314   & 0.263489 &RRc  \\
V16   &       182.5304722 &  18.5446111   &   16.490$\pm$0.0288   &      0.278$\pm$0.0412  & 0.260 &    0.530    & 0.469   & 0.371192 &RRc  \\
V17   &       182.5443611 &  18.5809167   &   17.113$\pm$0.0086   &      0.442$\pm$0.0126  & 2.515 &    0.486    & 0.358   & 0.375968 &RRc  \\
V20   &       182.5524722 &  18.5768611   &   13.436$\pm$0.0176   &      -                 & 2.534 &    0.169    & -       & 0.253247  &EA?  \\
V21   &       182.5214722 &  18.5561111   &   18.483$\pm$0.0222   &      0.517$\pm$0.0301  & 0.855 &    0.366    & 0.293  & 0.192828  &EW   \\
V22   &       182.5123611 &  18.5542500   &   17.365$\pm$0.0143   &      0.542$\pm$0.0193  & 1.062 &    0.415    & 0.315  & 0.539104  &E     \\
V23   &      182.5037778  & 18.5540833    &   16.677$\pm$0.0072   &      0.567$\pm$0.0097  & 1.464 &    0.155    & 0.154  & 0.474388  &?     \\
V24   &      182.5272222  & 18.5506111    &   16.991$\pm$0.0195   &      0.172$\pm$0.0323  & 0.480 &    0.151    & 0.170  & 0.290397  &EA    \\
V25   &      182.5358889  & 18.5472500    &   17.673$\pm$0.0211   &      0.532$\pm$0.0283  & 0.609 &    0.182    & 0.190  & 0.239222  &E     \\
V26   &      182.4740556  & 18.5471944    &   16.796$\pm$0.0090   &      0.538$\pm$0.0122  & 3.000 &    0.157    & 0.133  & 0.343250  &?     \\
V27   &      182.5318056  & 18.5468333    &   18.302$\pm$0.0541   &      0.443$\pm$0.0748  & 0.397 &    0.351    & 0.367  & 0.228866  &EW    \\
V28   &      182.5266944  & 18.5452500    &   16.439$\pm$0.0378   &      0.536$\pm$0.0508  & 0.155 &    0.343    & 0.302  & 0.295965  &RRc  \\
V29   &      182.5275278  & 18.5438056    &   17.003$\pm$0.0805   &      0.373$\pm$0.1220  & 0.093 &    0.515    & 0.452  & 0.304422  &RRc   \\
V30   &      182.5325833  & 18.5433056    &   17.393$\pm$0.0196   &      0.544$\pm$0.0253  & 0.358 &    0.131    & 0.110  & 0.323628  &E     \\
V31   &      182.5270278  & 18.5431667    &   16.547$\pm$0.0610   &            -           & 0.047 &    0.476    & -      & 0.282148  &RRc   \\
V32   &      182.5273889  & 18.5409167    &   16.361$\pm$0.0583   &      0.181$\pm$0.0890  & 0.137 &    0.478    & 0.388  & 0.296413  &RRc   \\
V33   &      182.5251389  & 18.5406944    &   17.289$\pm$0.0805   &      0.346$\pm$0.1124  & 0.137 &    0.616    & 0.553  & 0.296644  &RRc   \\
V34   &      182.5046667  & 18.5398333    &   19.419$\pm$0.0457   &      0.398$\pm$0.0656  & 1.250 &    0.722    & 0.322  & 0.438092  &RRc?  \\
V35   &      182.5277222  & 18.5395833    &   17.066$\pm$0.0505   &      0.456$\pm$0.0722  & 0.201 &    0.481    & 0.461  & 0.275419  &RRc   \\
V36   &      182.5205278  & 18.5374444    &   17.350$\pm$0.0196   &      0.062$\pm$0.0298  & 0.458 &    0.123    & 0.129  & 0.322817  &EA   \\
V37   &      182.5309444  & 18.5351111    &   15.653$\pm$0.0054   &      0.632$\pm$0.0086  & 0.527 &    0.079    & 0.062  & 0.291299  &EA     \\
V38   &      182.5162500  & 18.5323611    &   18.515$\pm$0.0266   &      0.471$\pm$0.0368  & 0.846 &    0.346    & 0.213  & 0.125619  &SX Phe \\
V39   &      182.5168611  & 18.5310000    &   18.081$\pm$0.0143   &      0.521$\pm$0.0192  & 0.886 &    0.329    & 0.293  & 0.104889  &EA    \\
V40   &      182.5325833  & 18.5304722    &   18.707$\pm$0.0340   &      0.432$\pm$0.0487  & 0.818 &    0.385    & 0.368  & 0.177438  &EW    \\
V41   &      182.5273056  & 18.5302222    &   17.512$\pm$0.0143   &      0.530$\pm$0.0202  & 0.753 &    0.167    & 0.146  & 0.540705  &EW     \\
V42   &      182.5347778  & 18.5301667    &   17.178$\pm$0.0108   &      0.096$\pm$0.0156  & 0.895 &    0.168    & 0.124  & 0.285490  &RRc   \\
V43   &      182.5132222  & 18.5245556    &   18.183$\pm$0.0126   &      0.517$\pm$0.0169  & 1.322 &    0.285    & 0.251  & 0.255195  &RRc?  \\
V44   &      182.5509167  &  18.4948333    &  18.505$\pm$0.0191  &       1.026$\pm$0.0239  & 3.209 &    0.268    & 0.168  & 0.232145  &EA?   \\ 
V45   &      182.5413333  &  18.5177778    &  18.920$\pm$0.0197  &       0.495$\pm$0.0266  & 1.730 &    0.297    & 0.311  & 0.313924  &?     \\ 
V46   &      182.5169722  &  18.5400278    &  18.420$\pm$0.0614  &       0.418$\pm$0.0896  & 0.559 &    0.502    & 0.497  & 0.476326  &EA    \\    
V47   &      182.5190000  &  18.5374167    &  16.982$\pm$0.0180  &       0.166$\pm$0.0300  & 0.525 &    0.063    & 0.128  & 0.424284  &EW   \\ 
\hline
\end{tabular}
\end{table*}
\begin{table*}
\caption{Membership probabilities of variable stars detected in the present work. Proper motion data of
variable stars have been taken from Wang et al. (2000) and Gaia astrometric mission. 
}
\tiny
\begin{tabular}{lccllllll}
\hline
ID   &   $\mu_{RA}$& $\mu_{Dec}$ &Prob.& ($\mu_{RA}$)gaia     &($\mu_{Dec}$)gaia & Prob.  & $P_{geom}$ & remark\\
     &    mas/yr   & mas/yr   &        &      mas/yr          &      mas/yr      &        &              & \\
    V1&       -    &     -    &     -  &     -1.996$\pm$0.162 & -2.123$\pm$0.121&  0.96    &        0.734  & 1 \\
    V2&      -3.22 &   -4.50  &  0.97  &     -1.780$\pm$0.164 & -2.272$\pm$0.108&  0.97    &        0.973  & 1 \\
    V3&      -5.22 &   -1.29  &  0.93  &     -1.694$\pm$0.170 & -1.884$\pm$0.126&  0.96    &        0.955  & 1 \\
    V4&      -5.40 &   -9.48  &  0.80  &     -1.311$\pm$0.171 & -1.990$\pm$0.105&  0.97    &        0.994  & 1 \\
    V6&      -3.67 &   -7.12  &  0.88  &     -1.733$\pm$0.152 & -2.001$\pm$0.100&  0.96    &        0.963  & 1 \\
    V7&       -    &     -    &     -  &     -5.333$\pm$0.394 & -2.910$\pm$0.271&  0.97    &        0.999  & 1 \\
    V8&       2.71 &    0.58  &  0.89  &     -0.826$\pm$0.359 & -1.274$\pm$0.249&  0.96    &        0.998  & 1 \\
   V10&      -2.74 &   -3.36  &  0.98  &     -1.574$\pm$0.153 & -2.131$\pm$0.099&  0.97    &        0.924  & 1 \\
   V11&      -3.57 &   -1.11  &  0.97  &    -12.302$\pm$2.635 & -3.221$\pm$1.332&  0.83    &        0.958  & 1 \\
   V12&      12.40 &  -14.95  &  0.71  &     -3.573$\pm$0.318 & -1.966$\pm$0.229&  0.95    &        0.999  & 1 \\
   V13&      -7.28 &  -12.71  &  0.70  &     -2.105$\pm$0.159 & -2.102$\pm$0.108&  0.96    &        0.992  & 1 \\
   V14&       2.71 &    0.58  &  0.89  &     -2.280$\pm$0.199 & -1.890$\pm$0.162&  0.96    &        0.998  & 1 \\
   V16&       1.66 &   -0.38  &  0.95  &      1.334$\pm$0.946 & -4.062$\pm$0.659&  0.94    &        0.995  & 1 \\
   V17&      -5.26 &   -6.11  &  0.89  &     -2.020$\pm$0.153 & -2.126$\pm$0.094&  0.96    &        0.488  & 1 \\
   V20&     -24.28 &   -1.05  &  0.00  &          -          &     -           &       -  &        0.481  & 0 \\
   V21&      -     &      -   &     -  &     -0.847$\pm$0.373 & -3.734$\pm$0.226&  0.97    &        0.941  & 1 \\
   V22&      -     &      -   &     -  &     -1.795$\pm$0.222 & -2.107$\pm$0.115&  0.97    &        0.909  & 1 \\
   V23&     -16.84 &  -13.95  &  0.00  &     -1.506$\pm$0.119 & -2.136$\pm$0.079&  0.97   &        0.827  & 0 \\
   V24&     -14.37 &   -8.29  &  0.71  &     -2.813$\pm$0.216 & -2.417$\pm$0.175&  0.96   &        0.981  & 1 \\
   V25&      -     &     -    &     -  &     -1.820$\pm$0.206 & -2.039$\pm$0.135&  0.96    &        0.970   & 1 \\
   V26&      -     &     -    &     -  &     -3.762$\pm$0.134 & -1.949$\pm$0.088&  0.95    &        0.272  & 0 \\
   V27&      -     &     -    &     -  &     -2.192$\pm$0.582 & -1.872$\pm$0.408&  0.96    &        0.987  & 1 \\
   V28&     -17.68 &   10.11  &  0.50  &     -1.802$\pm$0.166 & -2.552$\pm$0.118&  0.97   &        0.998  & 1 \\
   V29&      -     &     -    &     -  &    -17.932$\pm$2.546 &  2.378$\pm$2.101&    0.    &        0.999  & 1 \\
   V30&      -     &     -    &     -  &     -1.340$\pm$0.185 & -2.062$\pm$0.135&  0.97    &        0.990   & 1 \\
   V31&      -0.62 &   -4.22  &  0.90  &     -4.768$\pm$0.604 & -1.558$\pm$0.421&  0.97    &        1.000  & 1 \\
   V32&      -0.10 &  -25.17  &  0.01  &     -1.855$\pm$0.502 & -2.200$\pm$0.348&  0.97    &        0.998  & 1 \\
   V33&     -29.43 &   25.38  &  0.00  &             -       &      -          &       -  &        0.998  & 1 \\
   V34&      -     &    -     &     -  &     -1.846$\pm$1.579 & -3.189$\pm$0.553&  0.98    &        0.874  & 1/0 \\
   V35&      -     &    -     &     -  &            -        &              -  &     -     &        0.997  & 1 \\
   V36&      -     &    -     &     -  &     -0.578$\pm$0.412 & -3.463$\pm$0.269&  0.97    &        0.983  & 1 \\
   V37&      -0.08 &   -1.51  &  0.98  &     -1.693$\pm$0.078 & -2.025$\pm$0.071&  0.96    &        0.978  & 1 \\
   V38&      -     &    -     &     -  &     -4.146$\pm$0.498 & -1.296$\pm$0.333&  0.95    &        0.942  & 1 \\
   V39&      -     &    -     &     -  &     -1.796$\pm$0.257 & -2.183$\pm$0.173&  0.97    &        0.936  & 1 \\
   V40&      -     &    -     &     -  &     -4.803$\pm$0.589 & -1.424$\pm$0.439&  0.97    &        0.946  & 1 \\
   V41&      -     &    -     &     -  &     -2.129$\pm$0.193 & -2.123$\pm$0.145&  0.96    &        0.954  & 1 \\
   V42&      -0.33 &   -4.16  &  0.97  &           -        &          -    &    -      &        0.935  & 1 \\
   V43&      -     &    -     &     -  &     -1.572$\pm$0.2741& -1.685$\pm$0.185&  0.96    &        0.859  & 1/0 \\
   V44&      -     &    -     &     -  &           -       &         -    &  -        &        0.167  & 0 \\
   V45&      -     &    -     &     -  &           -       &         -    &  -        &        0.758  & 1 \\
   V46&      -     &    -     &     -  &     -1.309$\pm$0.5022& -3.606$\pm$0.500&  0.98    &        0.975  & 1 \\
   V47&      -0.50 &    5.36  &  0.53  &     -0.972$\pm$0.2271& -2.542$\pm$0.148&  0.97    &        0.978  & 1 \\
\hline         
\end{tabular}  
\end{table*}

\section{Identification of probable members of NGC 4147}
In order to understand the nature of variable candidates, it is necessary to find out their association to the globular cluster NGC 4147.  
We have used $V/(V-R)$ CMD and proper motion to identify probable members of the cluster as the sample of identified variables is contaminated by the field population.
The $V/(V-R)$ CMD of the cluster NGC 4147 discussed in the next section, clearly reveals contamination due to field star population. 

\begin{figure}
\includegraphics[width=9cm]{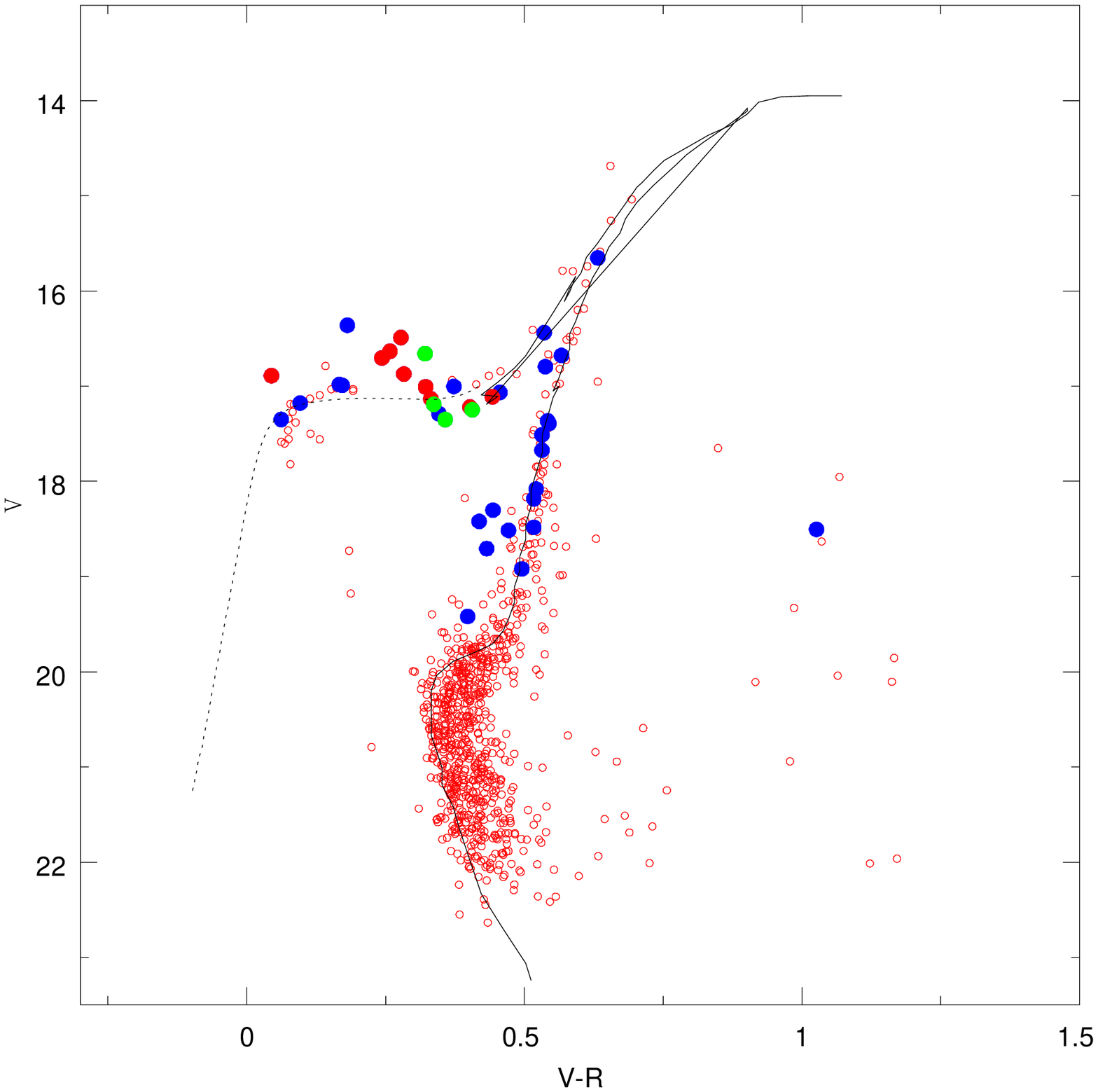}
\caption{$V/(V-R)$ CMD for stars in the region of NGC 4147. 
Variable stars identified in the present work are shown with larger filled circles. The continuous curve shows theoretical model by Girardi et al. (2002). The dotted line shows the ZAHB locus. Larger filled circles in green and red colour represent known RRab and RRc type variables, respectively. }
\end{figure}
\begin{figure*}
\includegraphics[width=16cm, height=10cm]{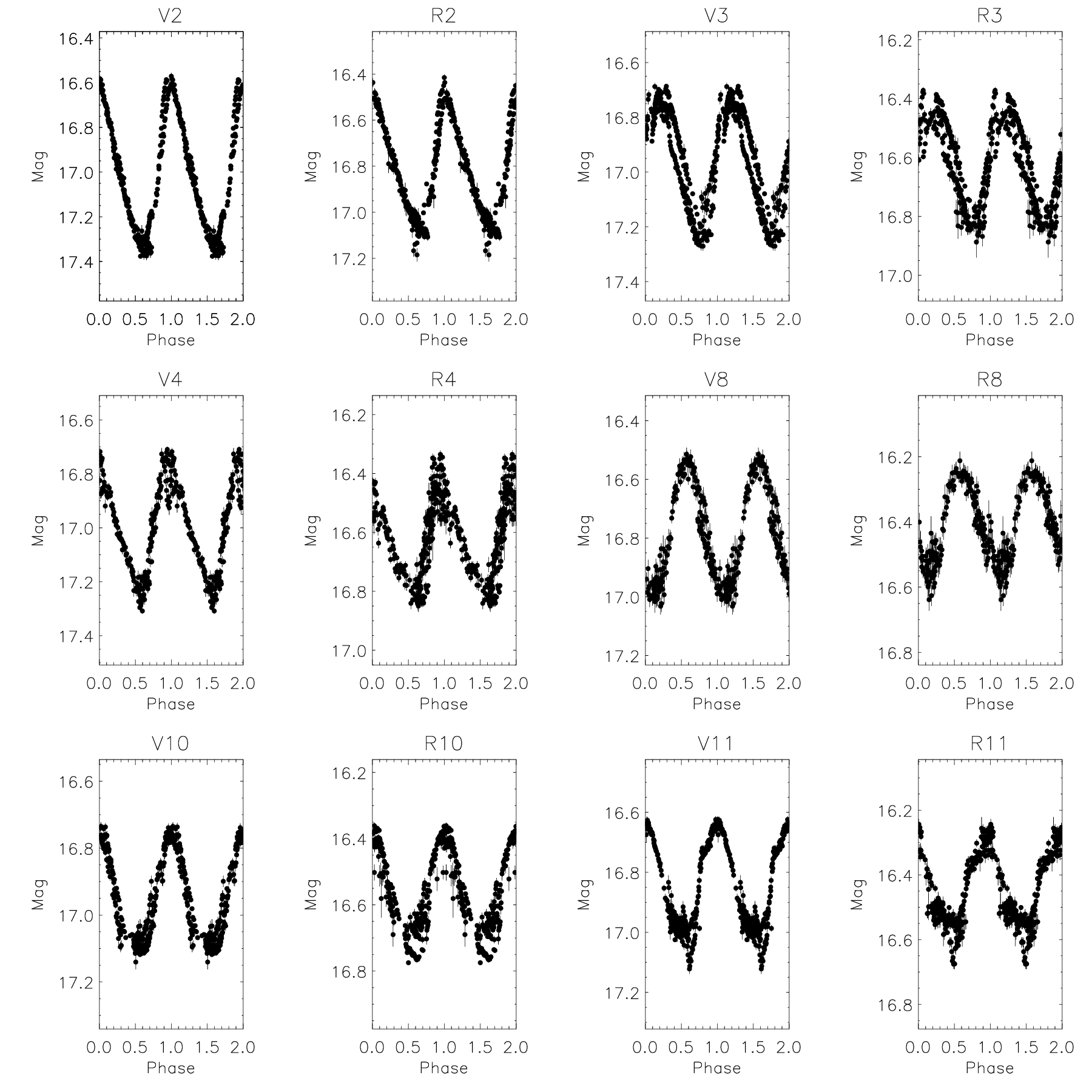}
\includegraphics[width=16cm, height=10cm]{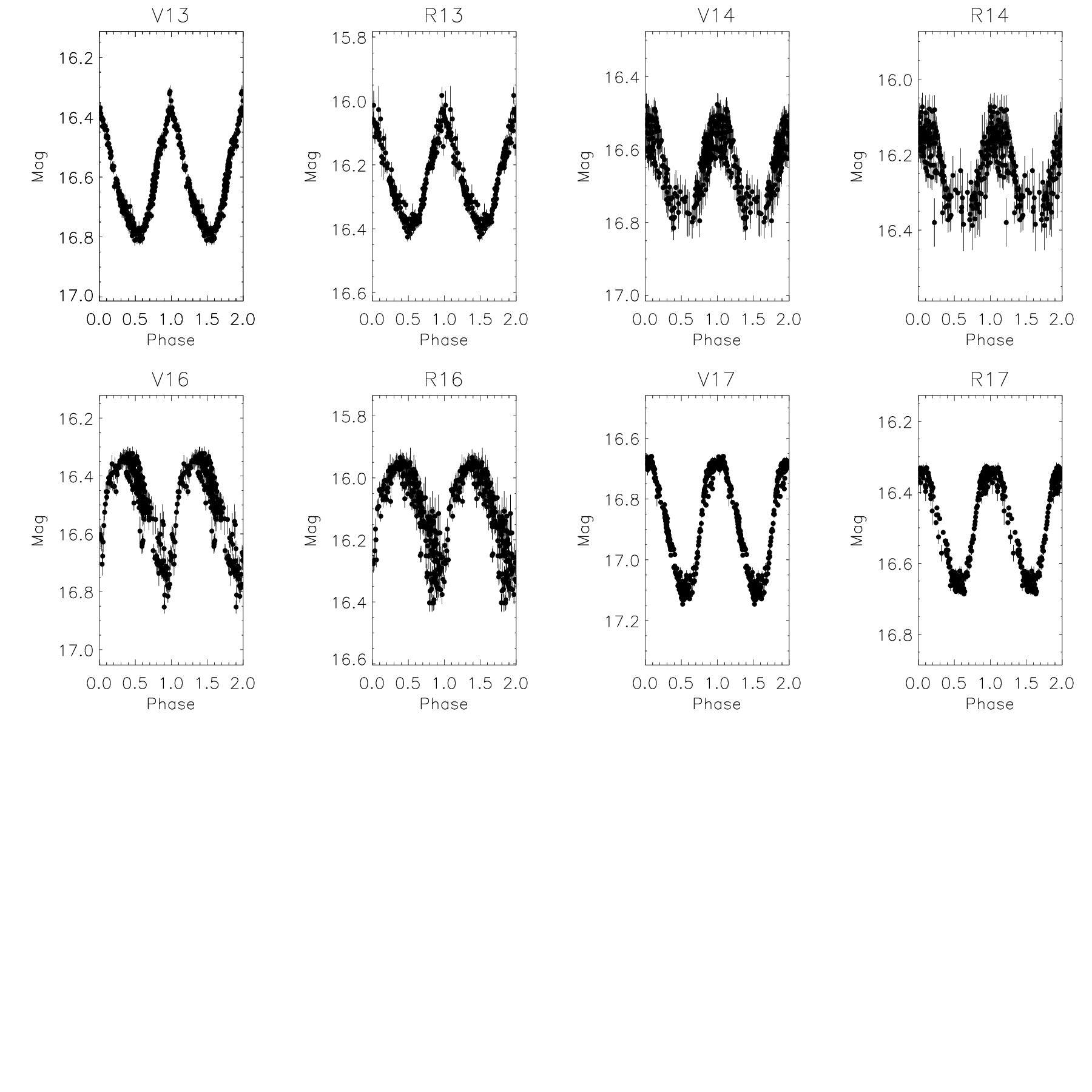}
\caption{The phased light curves of known variable stars in $V$ and $R$ bands. 
}
\end{figure*}
\subsection {$V/(V-R)$ colour-magnitude diagram}
Fig. 4 shows $V/(V-R)$ CMD for all the stars in the region of NGC 4147, which were common in  both $V$ and $R$ bands. In this Figure,  
larger filled circles represent variables identified in the present work. The continuous curve shows theoretical model by Girardi et al. (2002), while the dotted line represents the zero-age HB (ZAHB) locus which has been calculated using the Princeton-Goddard-PUC (PGPUC) stellar
evolutionary code based on Valcarce et al. (2012, 2013). PGPUC code is available for the chemical composition ranging from $Z = 1.60\times10^{-4}$ to 1.57$\times$10$^{-2}$(-2.25 $\lesssim$ $[Fe/H]$ $\lesssim$ -0.25), helium abundances from $Y$ = 0.23 to 0.37, and an alpha-element
enhancement of [$\alpha$/Fe] = 0.3. During calculation of the ZAHB locus
the value of mass, $Z$, $Y$ and [$\alpha$/Fe] has been taken as 0.8 $M_{\odot}$, 0.001, 0.23 and 0.3, respectively.
In Fig. 4, we have not plotted those stars which have large photometric errors \(>\) 0.02 mag up to $V$$\sim$ 18 mag. However, we have considered and  plotted those variable stars which have photometric errors \(>\) 0.02 mag.
In this  CMD, we could not plot three 
stars namely V12, V20 and V31 as their $(V-R) $ colours were not available.  The $V/(V-R) $ CMD clearly shows a 
well defined main-sequence, giant, supergiant, and HB.
From $V/(V-R)$ CMD, we have also noticed that there is no gap present in the HB.
The HB morphology of the cluster was discussed in detail by Stetson et al. (2005),  and 
they found that NGC 4147 has a predominantly blue HB which is different from Oosterhoff type I globular clusters.

An attempt has been made to determine age and distance of the cluster NGC 4147 by comparing present observations with the theoretical models of Girardi et al. (2002). 
The model of Girardi et al. (2002) for $Z$ (relative metal abundance) = 0.001 and $Y$ (relative helium abundance)=0.23 gives the best fit to the observed distribution of stars in the CMD of NGC 4147.  
The reddening towards the cluster region was estimated along with the age and distance modulus by comparing present observations with 
theoretical models.
The distance modulus $(V-M_{V})$ and age of the cluster are thus obtained as 16.40 mag and 12.58 gyrs, respectively. The derived age of NGC 4147 is consistent with the age of other globular clusters. Using distance modulus the distance to the cluster was calculated to be 17.49 kpc.  The value of reddening $E(V-R)$ towards the cluster region estimated as 0.04 mag which further yielded $E(B-V)$ ($=E(V-R)/0.65$) reddening value as 0.06 mag.  
The metallicity $[Fe/H]$ of stars can be 
generally calculated using the following relation $[Fe/H]=\log(Z/X)-\log(Z/X)_{\odot}$, where $Z/X$ is the metal 
to hydrogen ratio (Harris 1996: 2010 Edition), Smith 1995). The above conversion is used if the elemental distribution is 
assumed to follow the solar abundances.  The metals in stars are often grouped into two 
categories: $\alpha$ elements and Fe elements.  The old metal poor stars are typically $\alpha$ enhanced 
in the halo and bulge of our Galaxy ([Fe/H] $<$ $-0.6$ and [$\alpha$/Fe] $\sim$0.3-0.4) (e.g. C{\'a}ceres \& Catelan 2008 and references therein).  Therefore, to determine the relative iron abundance with respect to the 
solar iron abundance for any [$\alpha$/Fe] value we have used the relation given by Valcarce et al. (2012, 2013). Thus, the value of metallicity $[Fe/H]$ with [$\alpha$/Fe]=0.3 comes out be $-1.47$. If we consider the value of [$\alpha$/Fe]=0.4, the metallicity $[Fe/H]$ comes out to be  $-1.57$.  

All the present variable candidates except V44 are found to be located on the HB, red giant branch and blue stragglers region, and
these could be probable members of the cluster.
Due to lack of $R$ band data the membership of newly identified variables V20 and
V31 can not be determined.

\begin{figure}
\includegraphics[height=9cm]{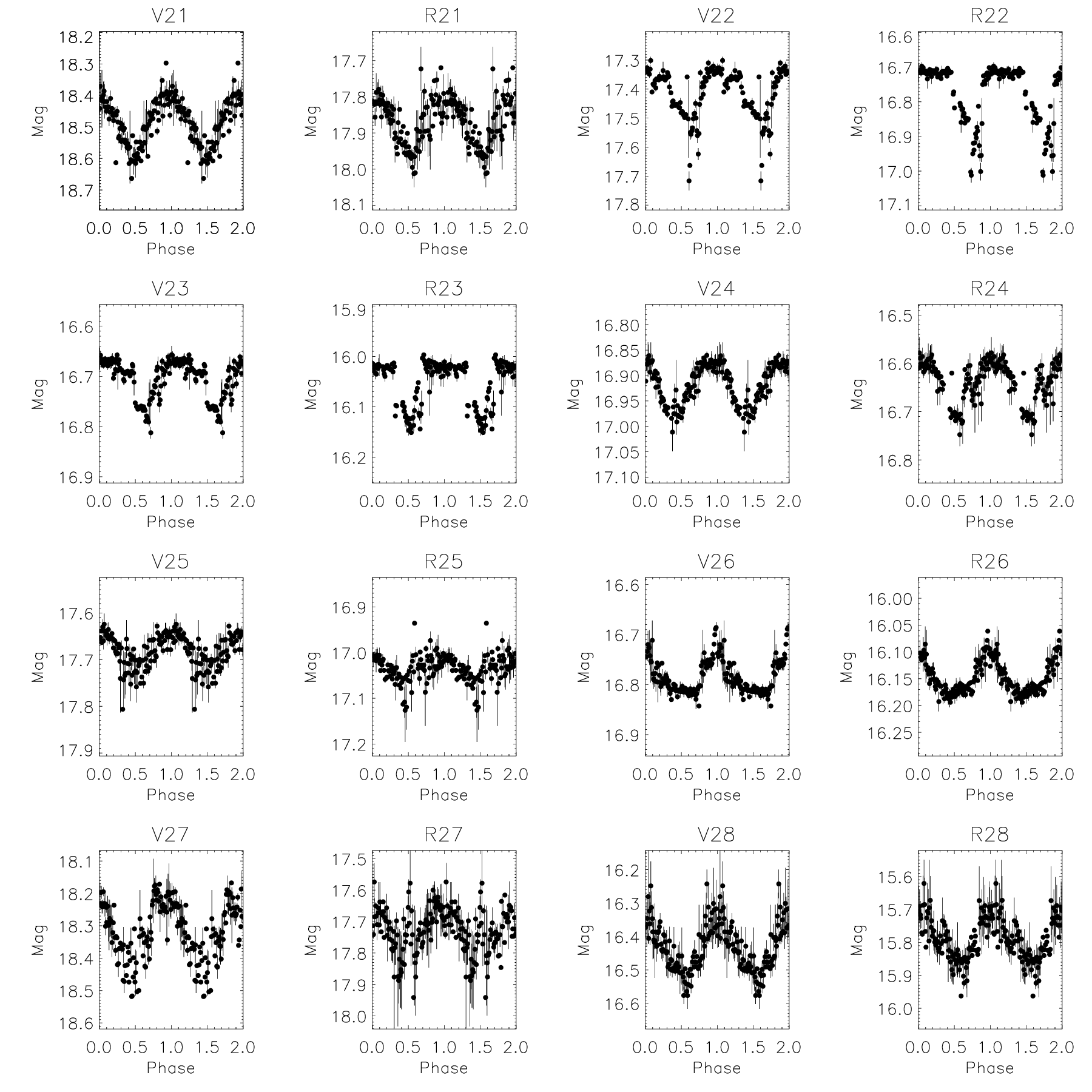}
\includegraphics[height=9cm]{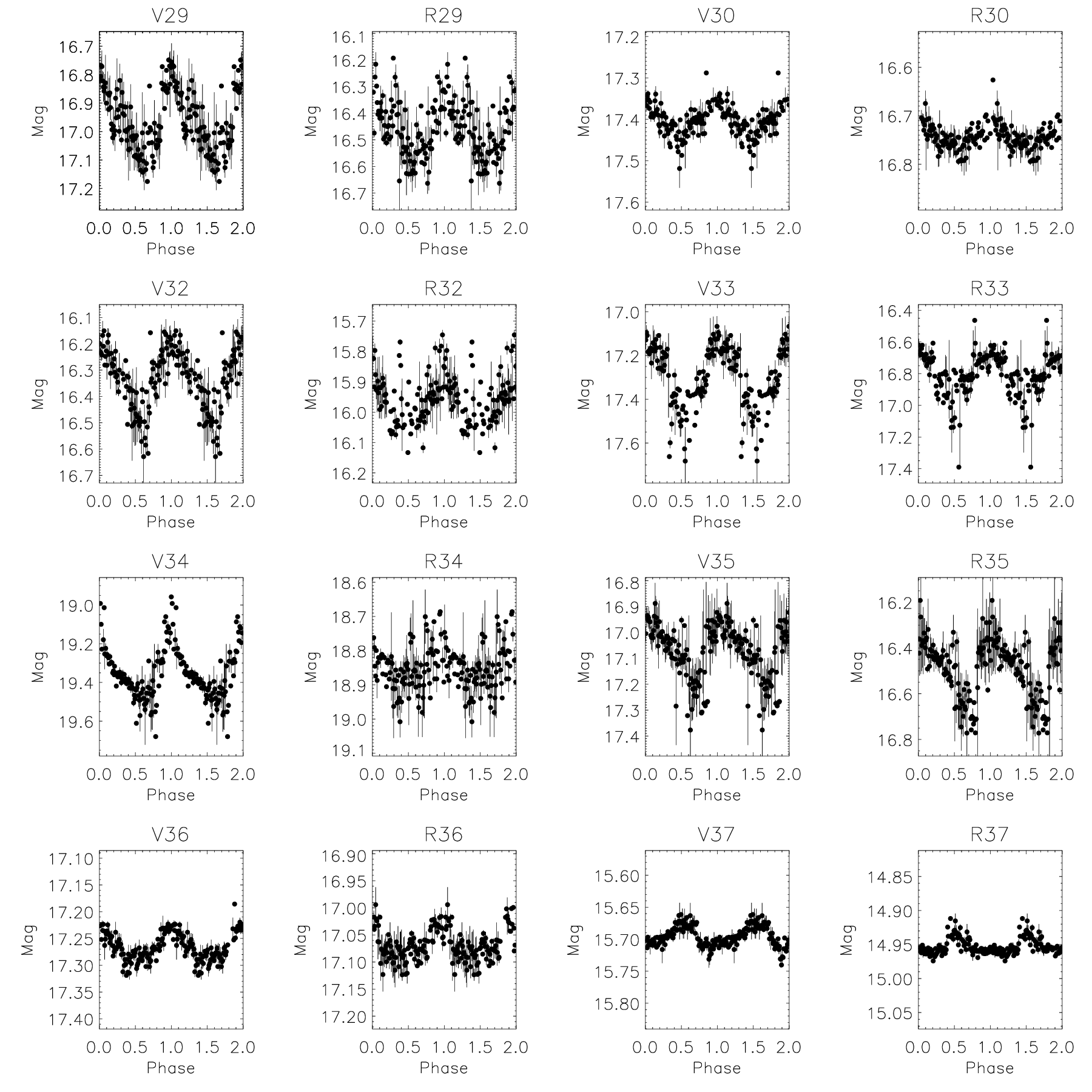}
\caption{The phased light curves of newly identified variable stars in $V$ and $R$ bands. 
}
\end{figure}
\setcounter{figure}{5}
\begin{figure}
\includegraphics[height=9cm]{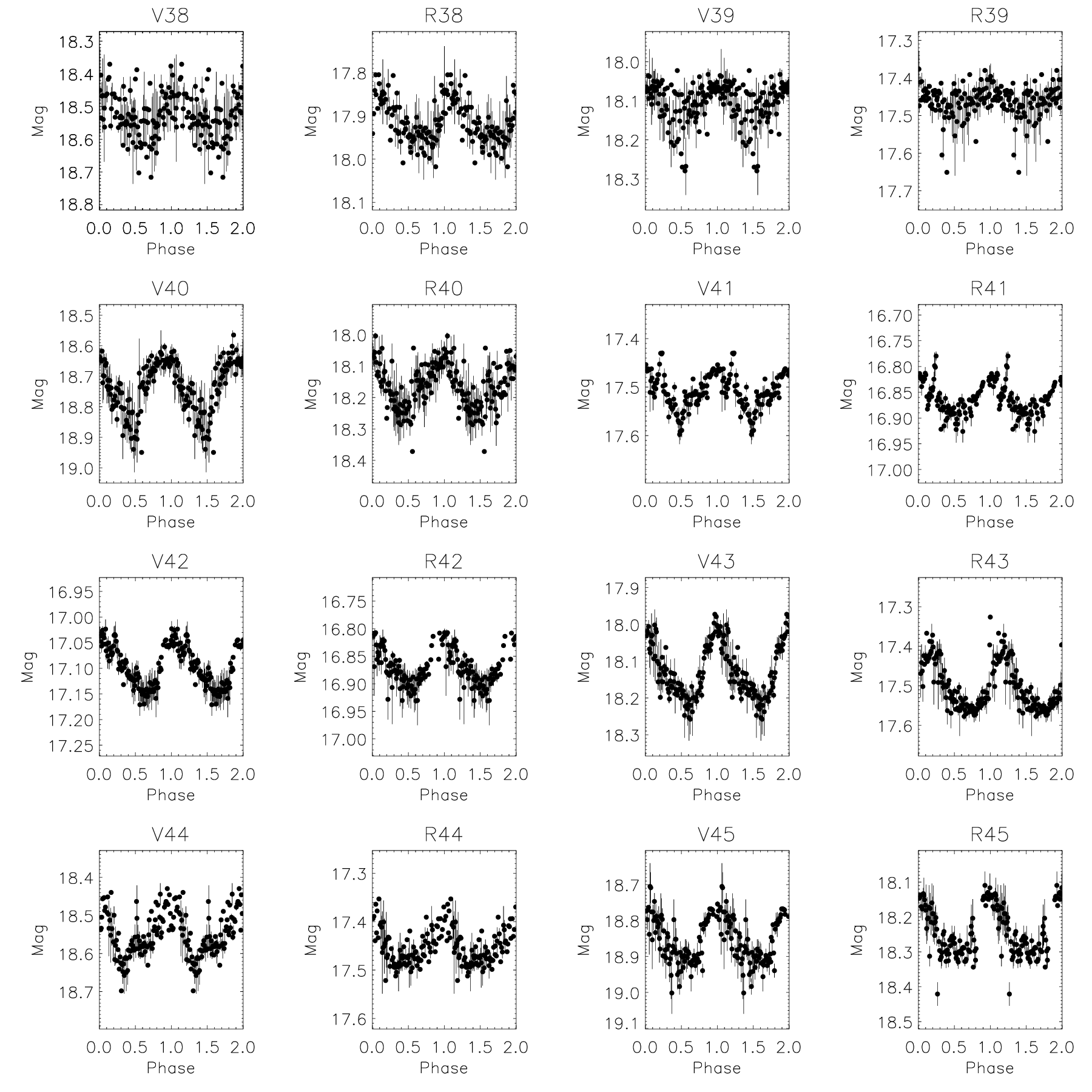}
\includegraphics[height=9cm]{un4_ref.ps}
\caption{Continued. 
}
\end{figure}

\subsection{Membership probability based on  Proper motions and geometric distribution}
The proper motion data by Wang et al. (2000) have also been used to determine the membership of identified
variables. Wang et al. (2000) have estimated proper motion of stars having magnitudes B $\lesssim$17.0 mag using
three epoch data. As per their criterion a star having membership probability $\le$0.67
can be regarded as a non-member by them.
Wang et al. (2000) found that there are quite a few stars with large membership probabilities located above the HB. From their location in the $V/(V-R)$ CMD of NGC 4147, they
suggested the existence of a second HB, and it might be due to their binary nature. 
For the larger dispersion in the member star sequence Wang et al. (2000) suggested that there could be crowding problems near the cluster center that reduced the accuracy of the photometric data. 

Present data of variables have been cross-matched with Wang et al. (2000) data within the 1 arcsec matching radius and 
23 stars were found common between these two data sets. 
The known variables V2, V3, V4, V6, V8, V10, V11, V12, V14, V16 and V17 detected in the present work have membership probabilities $\ge$ 0.70, hence these could be the members of the cluster. Wang et al. (2000) considered V13 as non-member of
the cluster as its membership probability has been found to be 0.70.
Proper motion data of known variable star V1 is not known.
Only 10 newly identified variables have proper motion data, of which
four stars V24, V31, V37 and V42 have membership probabilities $\ge$ 0.70. For stars V20, V23, V28, V32, V33 and V47 membership probability was found to be either zero or $\le$ 0.70. The proper motions of cross-matched stars along with their membership probability are listed in Table 3.

We have also used proper motion values of variable stars taken from Gaia astrometric mission (Gaia Collaboration et al. 2018). The Gaia astrometric mission has been launched in 2013 
to measure positions, parallaxes, proper motions and photometry to obtain physical parameters for millions of stars. 
There are 36 variable  stars which have proper motion values. For calculation of membership probabilities using the proper motion data of these
36 stars, we tried  parametric approach which is commonly used 
as described in Vasilevskis, Klemola \& Preston (1958). 
This approach assumes a normal bivariate distribution for proper motion values for both field and cluster stars. Distribution of proper motion values for field stars around our cluster followed no standard form and therefore, we used the non-parametric approach as described in 
 Cabrera-Cano \& Alfaro (1990) through its implementation Clusterix (http://clusterix.cerit-sc.cz/). 
It iteratively improves empirically determined cluster and field stars distributions, without 
any assumption about their shape and also uses positional data as supplementary information. The proper motion 
cutoff was set to 10 mas/yr and radius of the cluster to 6.5 arcmins. Table 3 lists proper motion data and membership probability of 36 variable stars.
Table 3 shows that all these stars excluding V29 have membership probability more than 80\%.

Rozyczka et al. (2017) discussed membership probabilities of the variable stars
 based on their proper motions, spatial distribution and CMD location.
They allow for the fact that stars with proper motion
membership probabilities less than 70\% might still be cluster
members.  In the case of  M22, they proposed that if the
proper motion probability of the star is less than 70\%, but if it has geometric probability $>$ 99\% with condition that if the star's CMD location is  
 appropriate
for its variability type, and its
distance from the cluster center is less than 3.3 arcmin
(which is the half light radius for M22), the star could be a
 member of the cluster.    
This could be situation for V28, V32, V33 and V47 in NGC 4147. 
These stars are located in the crowded central region of the
cluster within the half light radius $\lesssim$0.48 arcmin, where it might have been difficult to derive accurate
proper motions and distances. On the other hand, V20 and V23
both have membership probabilities less than 70\% and are located
much further from the cluster center.  They are probably field
stars.
The geometric membership probabilities of stars have been calculated  using the relation $P_{geom} = 1-\pi r^2 /S$ given by Rozyczka et al. (2017), where $r$ 
is the radial distance from the cluster center in arcsec and $S$ is the size of the field of view in arcsec$^2$. The value of $S$ is taken as 139726.644 arcsec$^2$. The determined
geometric membership probabilities of variable stars are listed in Table 3.

To assign membership status of the stars when no proper motions are available,  
Rozyczka et al. (2017) suggested another membership criteria based on geometric probabilities $>$70\%, CMD
location and distance from the cluster center.
Using this criteria stars V21, V22, V25, V26, V27, V29, V30, V34, V35, V36, V38, V39, V40, V41, V43, and V46 could be probable members of the cluster. 
Two stars V26 and V44 which are located outside the half light radius and have geometric probabilities $<$70\% can be considered as field stars. 
Thus, the membership status of the each star decided on the proper motion, CMD location and geometric membership probabilities is given in Table 3.
In Table 3, the member stars are flagged as `1', whereas non-members as `0'.  

\begin{figure*}
\hbox{
\includegraphics[height=7cm,angle=270]{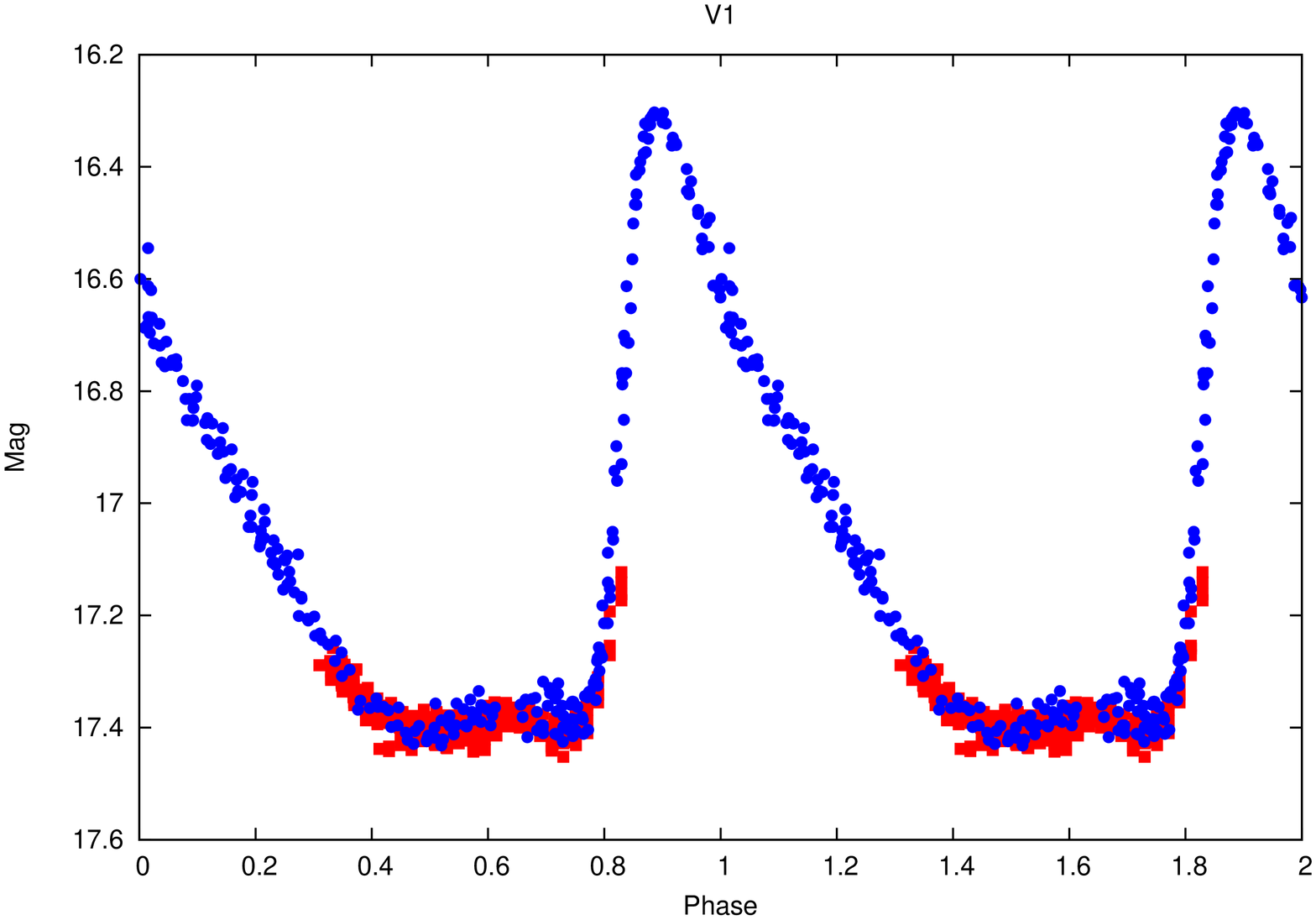}
\includegraphics[height=7cm,angle=270]{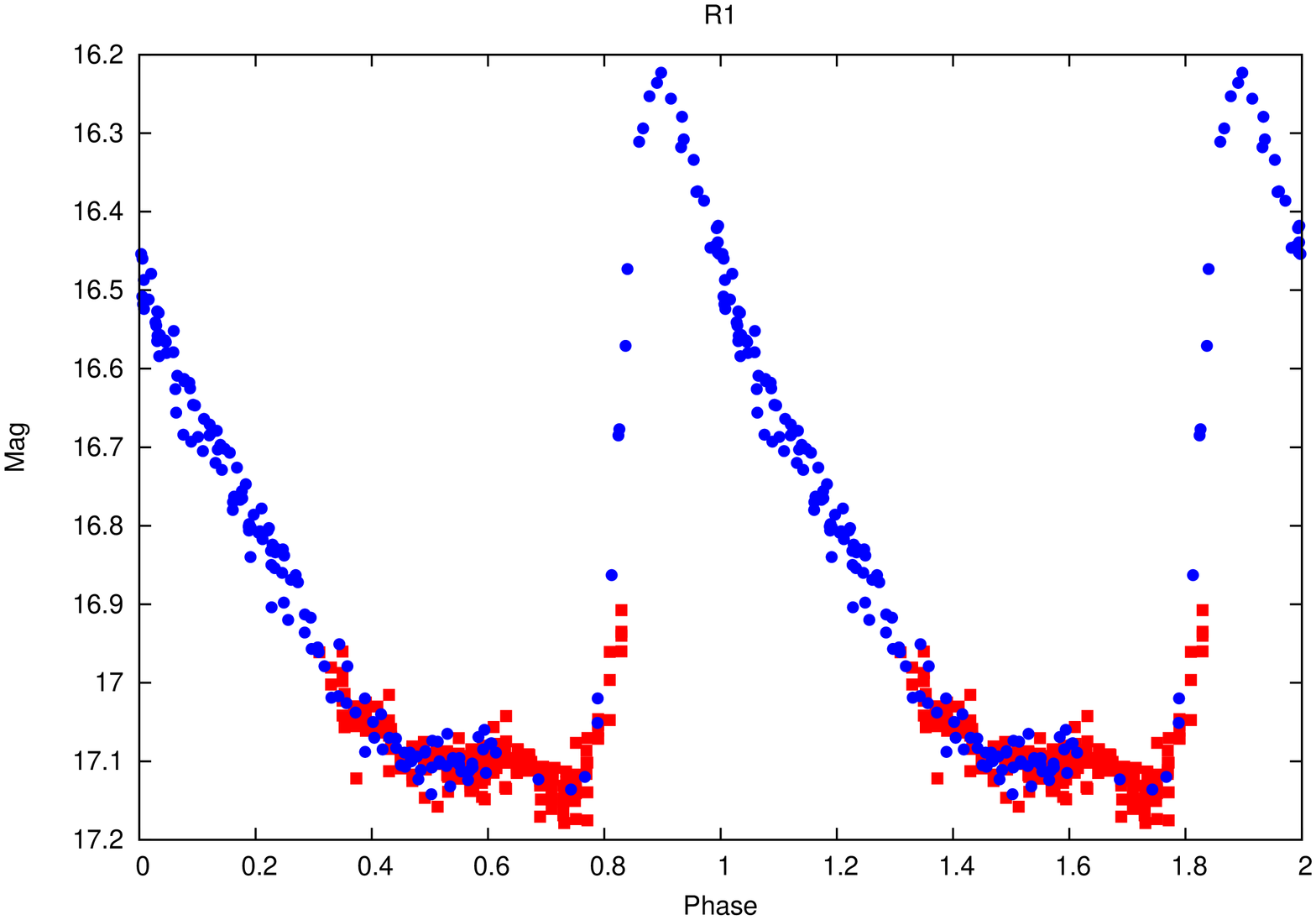}
}
\hbox{
\includegraphics[height=7cm,angle=270]{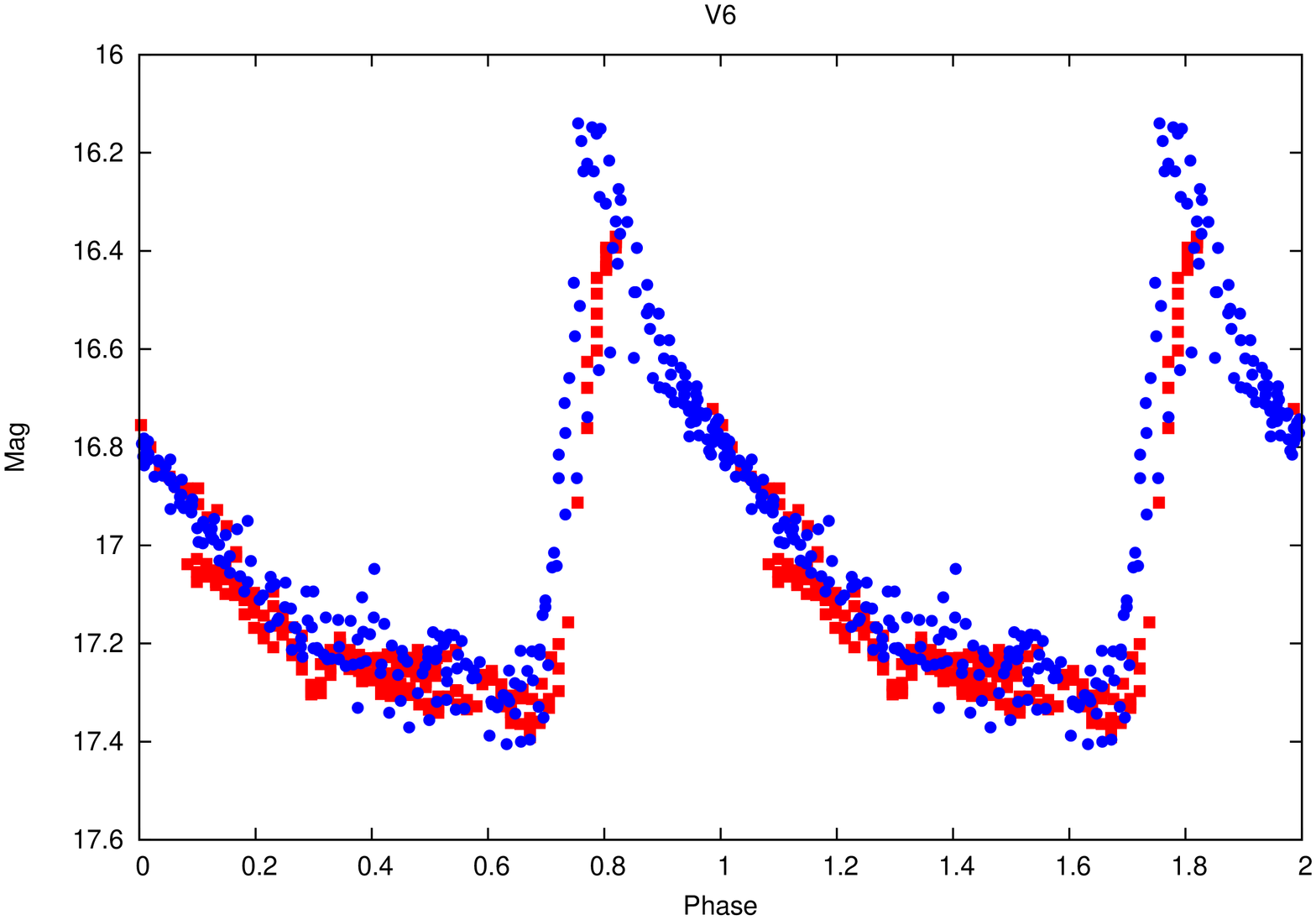}
\includegraphics[height=7cm,angle=270]{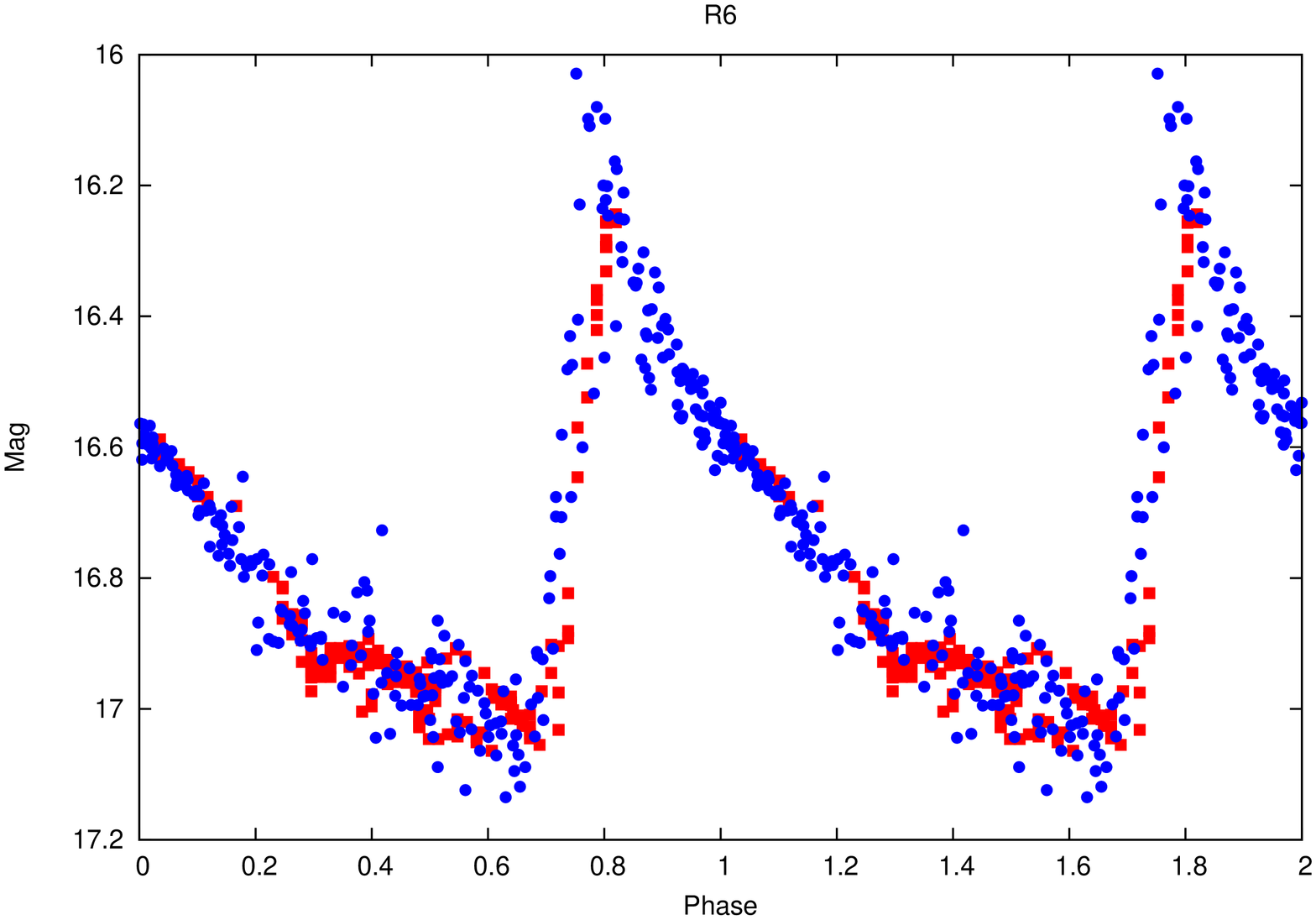}
}
\hbox{
\includegraphics[height=7cm,angle=270]{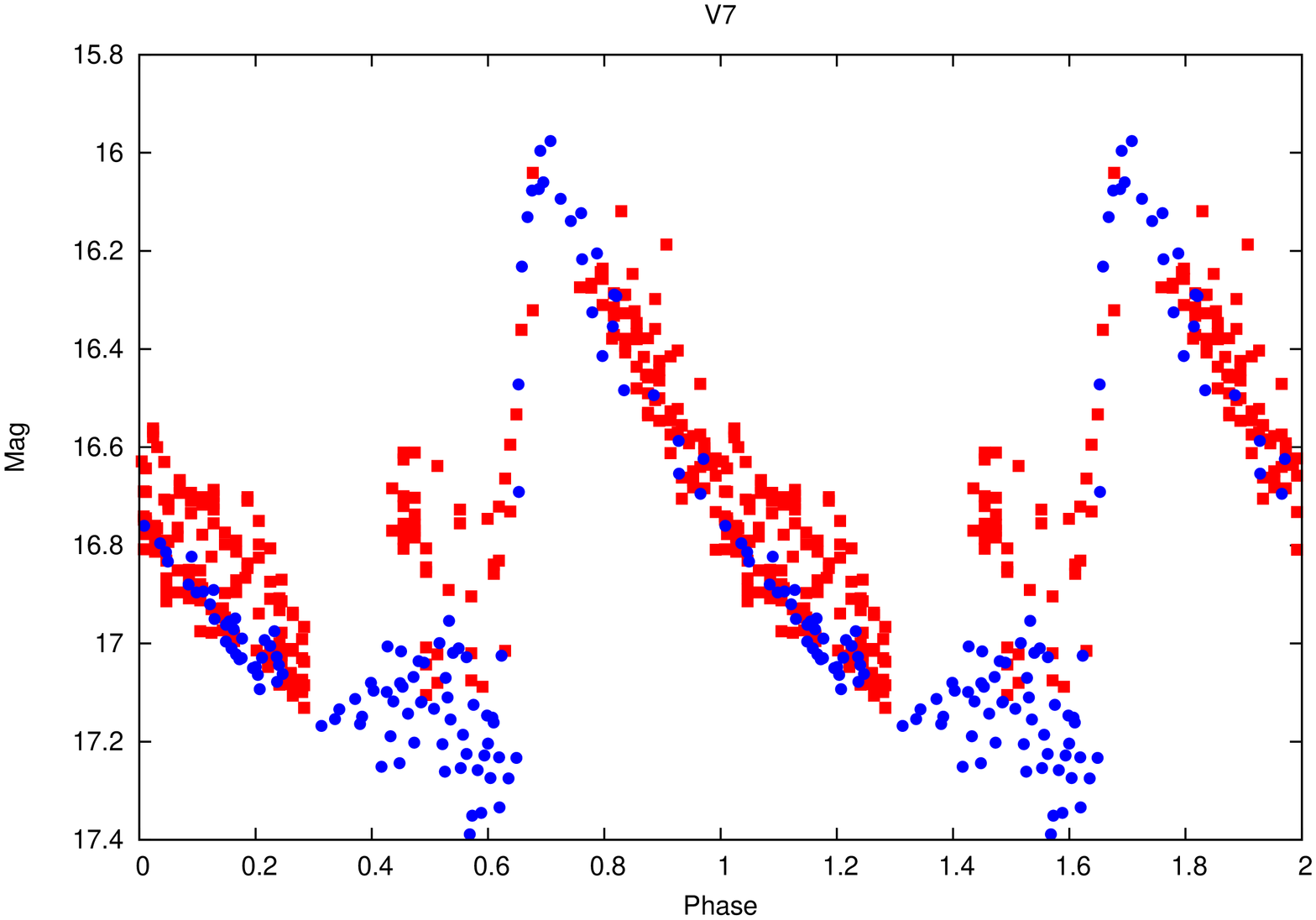}
\includegraphics[height=7cm,angle=270]{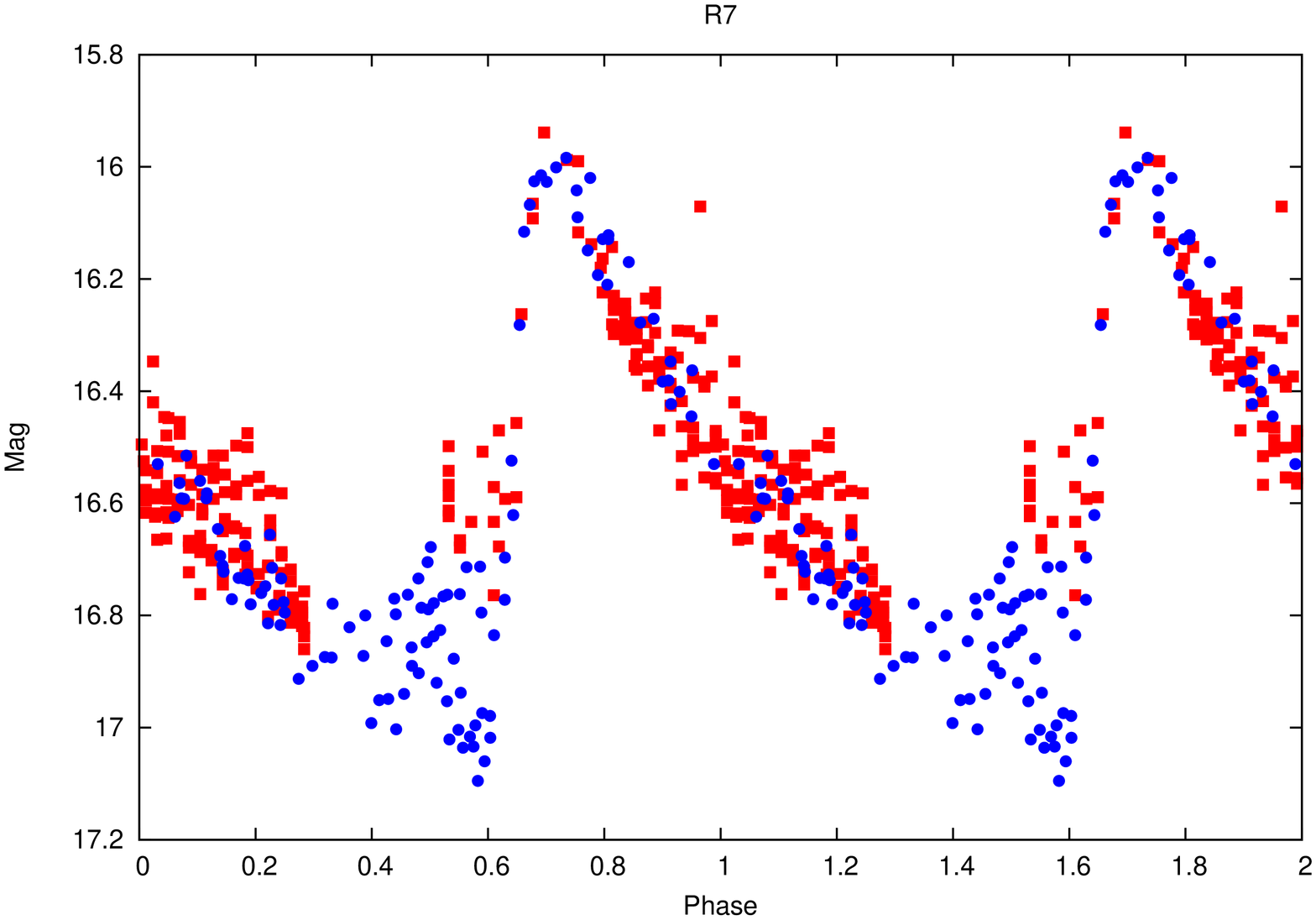}
}
\hbox{
\includegraphics[height=7cm,angle=270]{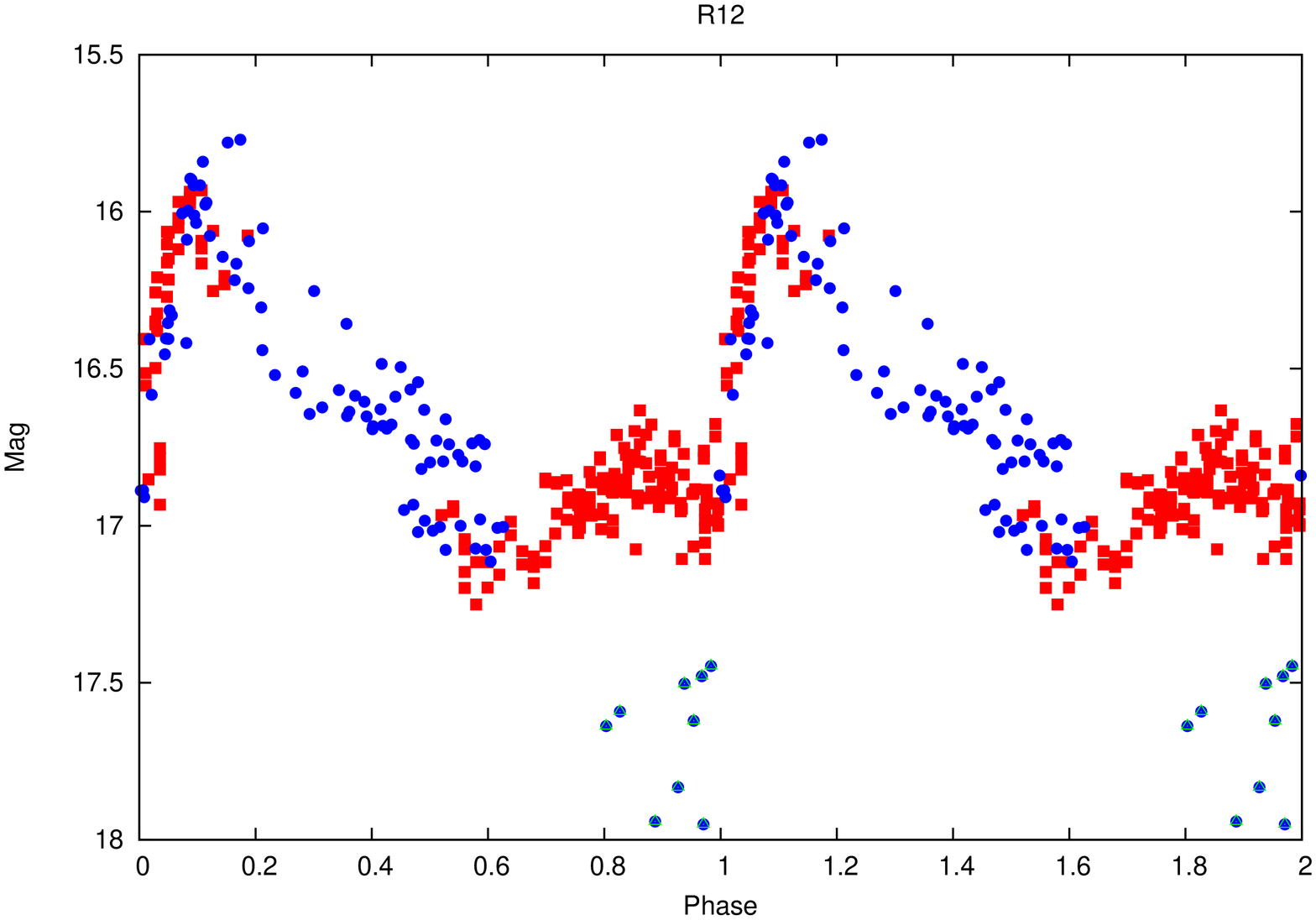}
}
\caption{The phased light curves of V1, V6, V7 and V12 in $V$ and $R$ band. The filled squares show present data while filled circles 
represent data of Arellano Ferro et al. (2004). In the case of V12 isolated data points are displayed by asterisk symbol. 
}
\end{figure*}

\begin{figure*}
\centering
\hbox{
\includegraphics[height=9cm,angle=270]{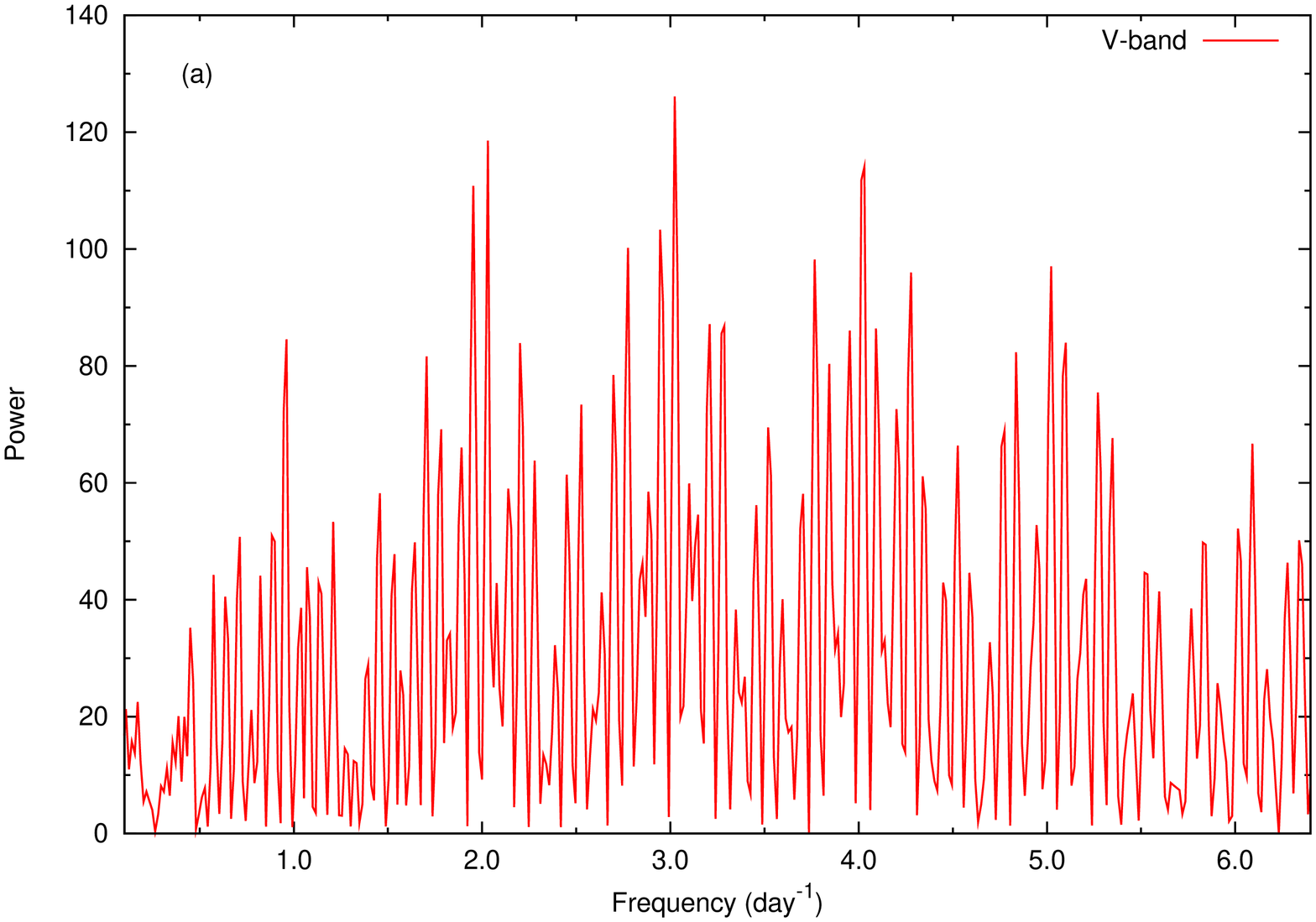}
\includegraphics[height=9cm,angle=270]{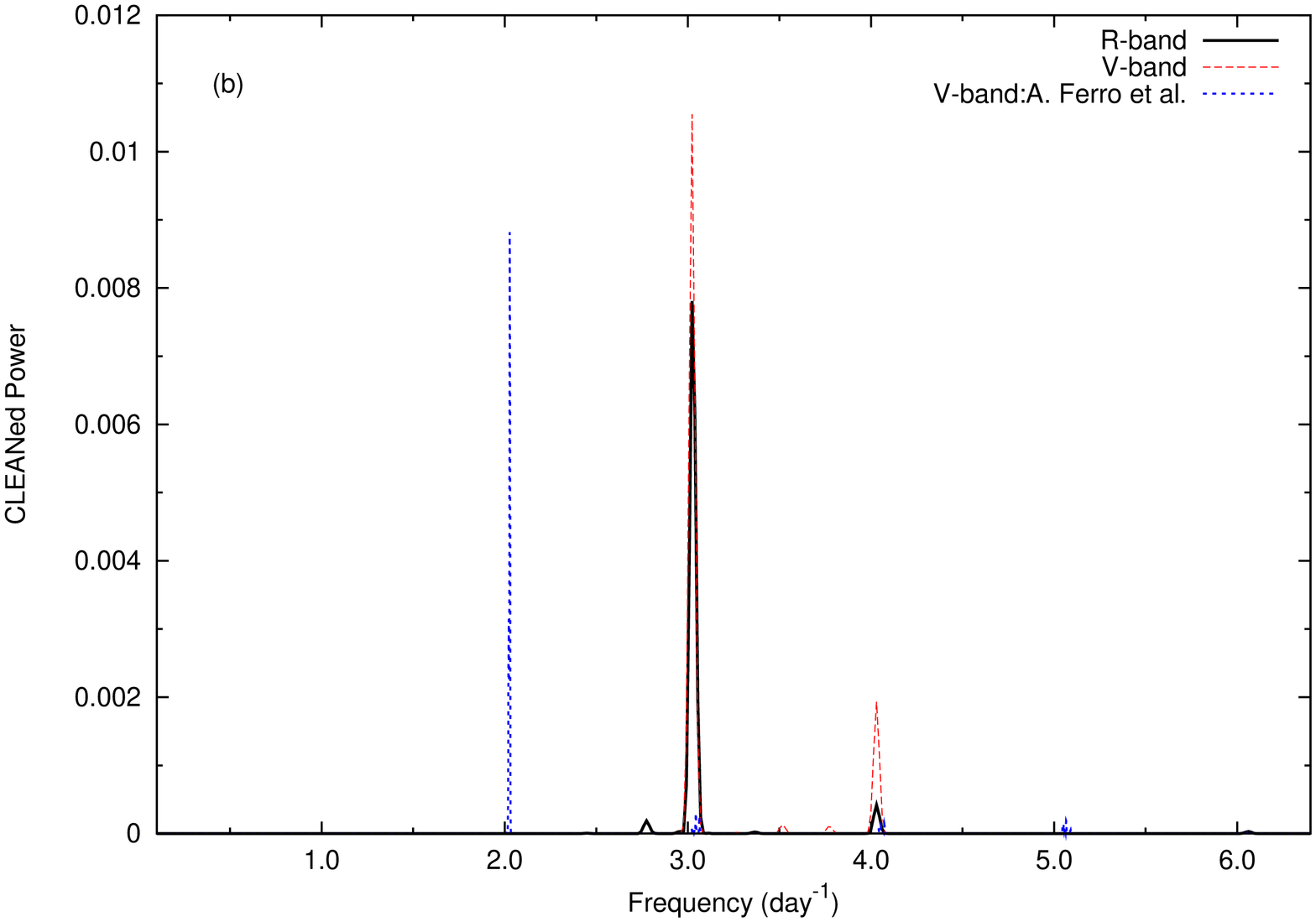}}
\hbox{
\includegraphics[height=9cm,angle=270]{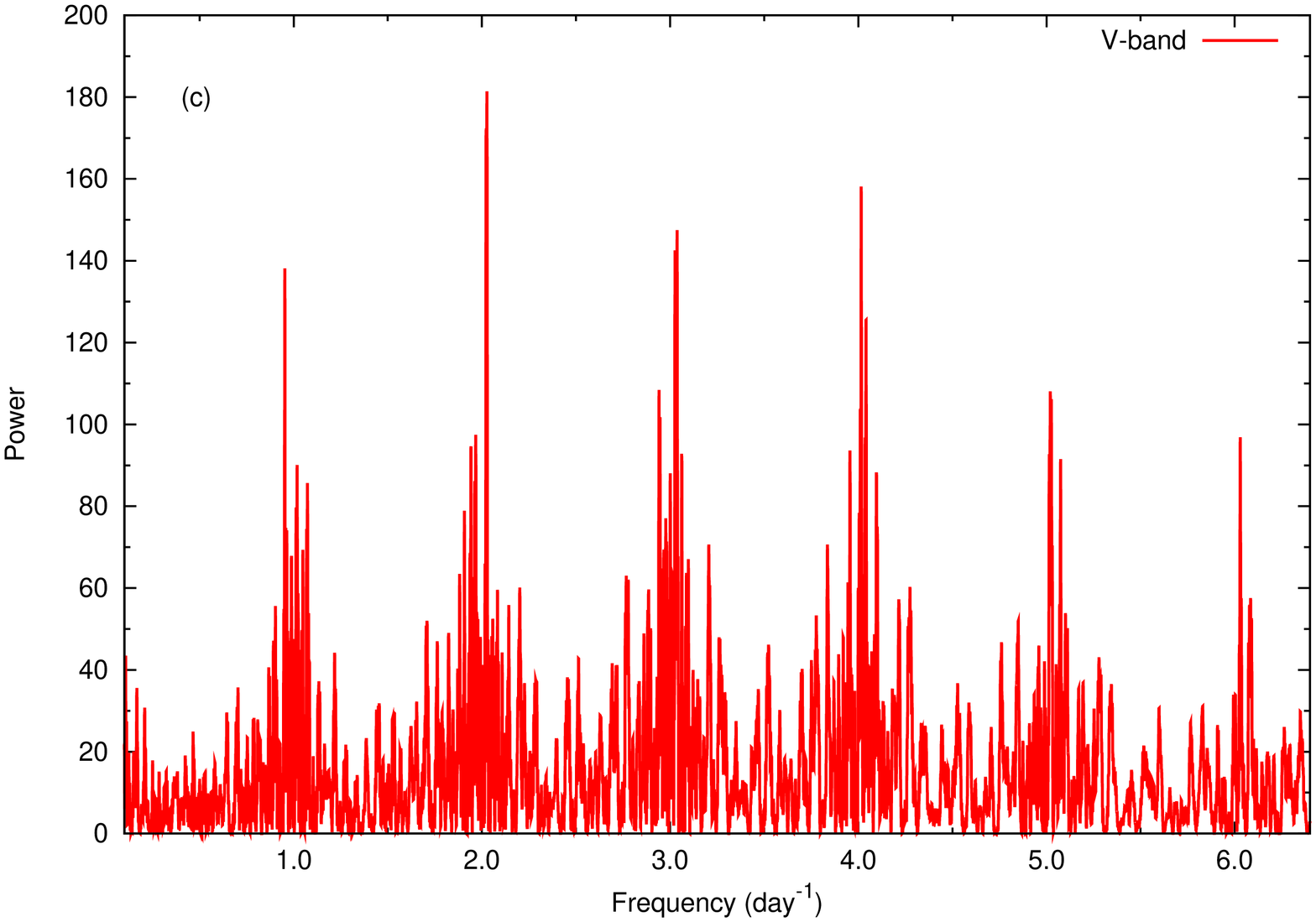}
\includegraphics[height=9cm,angle=270]{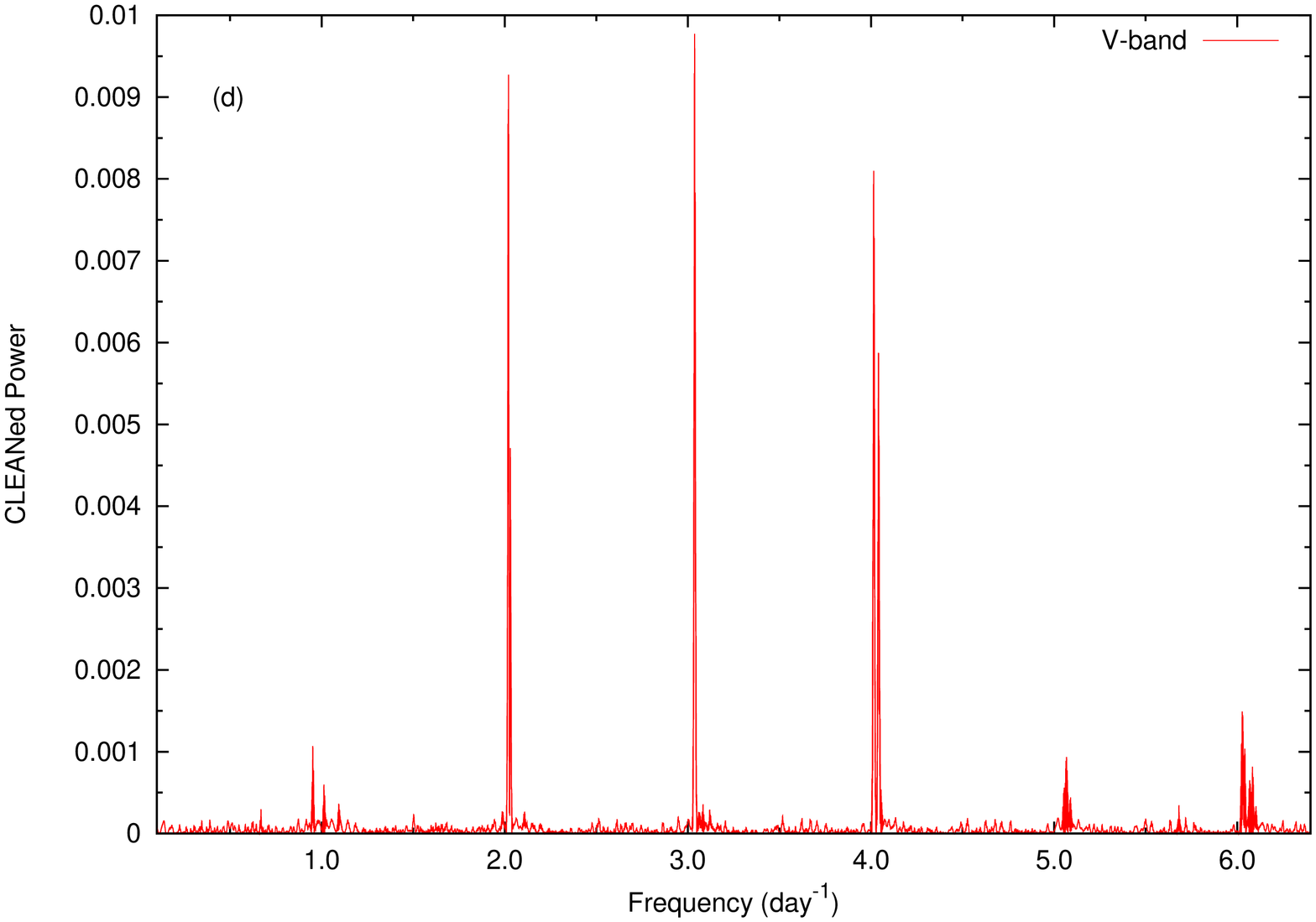}
}
\caption{Power spectrum of star V2. Plot (a) represents power spectrum obtained using LS algorithm for present $V$-band data while plot (b) shows power spectrum obtained from CLEANed algorithm. Small dashed line shows power spectrum obtained using $V$ band data of Arellano Ferro et al. (2004).
Long dashed and continuous curve represents present data in $V$ and $R$ band, respectively.
Plot (c) shows power spectrum of V2 obtained using LS algorithm in combination with previous data (Arellano Ferro et al. 2004), whereas (d) displays power spectrum obtained from CLEANed algorithm.
}
\end{figure*}

\section{Period determination}
We have used the Lomb-Scargle (LS) periodogram (Lomb 1976, Scargle 1982) to determine the most likely periods of the variable stars. This periodogram gives better periods even if the data are taken at irregular intervals.  
Derived periods from LS periodogram were further confirmed using the NASA exoplanet archive periodogram service\footnote{https://exoplanetarchive.ipac.caltech.edu/applications/Periodogram/}. 
Light curves of variable stars were folded using these derived periods. The phased light curves were visually inspected and we opted period which gives the best folded light curve. 
Figs. 5 and 6 display the phased light curves of known and newly identified variable stars, respectively.  The $R$ band light curves of variables have been phased with the period obtained from their $V$ band light curves.  
For better representation of the phased light curves for 28 newly discovered variable stars, we have  folded their light curves with the phase bin of 0.01. 

We could not determine the period of star V1 on the basis of present data due to
our short observational baseline.
We have combined present data of star V1 with  that of  Arellano Ferro et al. (2004) to determine its period.
A period of 0.50039 days for V1 was determined using LS periodogram. There was systematic offset  of 0.1 mag between present and  Arellano Ferro et al. (2004) data. We brought present magnitudes to the magnitudes of Arellano Ferro et al. (2004) by applying the offset. The same procedure has been adopted in the case of variables V6, V7 and V12 to determine their periods.
The phased light curves of V1, V6, V7 and V12 in both $V$ and $R$ bands are plotted in Fig. 7. 
The periods and amplitudes in $V$ and $R$ bands of 42 variable stars are given in Table 2. 

\section{RR Lyrae variables}
\subsection{Known}
The present study consists of 14 known variables which have already been discovered and studied. The studies of the cluster NGC 4147  by Davis (1917) and Baade (1930) discovered one and three variable stars, respectively.
The photometric study by Sandage \& Walker (1955)
reported 10 variable stars in the cluster region.
Newburn (1957) studied variable stars and discovered three variable candidates. A study on variable stars in NGC 4147 was presented by Mannino (1957).
Clement (1997) provided a list of 17 RR Lyrae stars in the region of NGC 4147. 
 Arellano Ferro et al. (2004) reported results of $V$ and $R$ band CCD photometry of the known RR Lyrae stars in NGC 4147 and derived significantly improved periods and discovered one new variable V18. 
 The periodicities of most variables were revised and new ephemerides were calculated by Arellano Ferro et al. (2004).
They have detected Blazhko effect in two stars namely V2 and V6 and found three variables V5, V9 and V15 as non-variables which had been previously reported as variables.
The variable V18 discovered by them 
 has period of $\sim$0.49205 days and amplitude of $\sim$0.15 mag. The physical parameters of RR Lyrae variables were also estimated by them.
Stetson et al. (2005) presented a detailed study on photometry and astrometry of the cluster NGC 4147. They reanalyzed five exposures of the cluster obtained with WFPC2 on the Hubble Space Telescope and presented calibrated CMDs and colour-colour  diagrams. They also found morphological properties which were generally
consistent with the previous published works. Star V18 discovered by Arellano Ferro et al. (2004) was
found non variable by Stetson et al. (2005). 
They have also detected one new variable V19 and characterized it as an RR Lyrae star.

The previously known RR Lyrae variables in the present work have periods in the range from $\sim$0.26 to $\sim$0.60 days and their amplitude of brightness variation in the $V$ band ranges
from $\sim$0.34 to $\sim$1.17 mag. Their amplitudes, periods and location in the $V/(V-R)$ CMD are in agreement with that of RR Lyrae-type stars.
The known RR Lyrae variables have already been classified as RRab and RRc type on the basis of their variability characteristics in earlier works. The RRab variables V1, V2, V6, and V7 are varying with larger 
$V$ amplitudes in the range of $\sim$1.02 to  $\sim$1.17 mag.  The periods of these RRab stars vary from $\sim$0.49 to $\sim$0.61 days.  All known RRc variables namely V3, V4, V8, V10, V11, V14, V16, and V17 have amplitudes in the range of $\sim$0.34 to $\sim$0.71 mag, while their brightness varies with the periods between $\sim$0.26 to $\sim$0.40 days. The periods of cataloged variables V3, V4, V6, V7 V8, V10, V11, V12, V13, and V17 are revised and found to be in agreement with previously determined periods. There are two stars V2, and V14 whose periods derived using the present data are different from the periods reported in the previous studies. 
 We could not detect brightness variation in stars V5, V9, V15, and V18 using present data.
This confirms the conclusion reached by Stetson et al. (2005) and that these stars are no longer considered to be variables.
Star V19 could not be detected in the present work because it was outside of the present field of view.

Now, we would like to discuss those known variables whose periods,
amplitudes and shape of light curves have been found to be slightly different from the previous studies. 

In the case of star V2 the derived period of 0.335070 days from the present data is not consistent with that of earlier derived period (Arellano Ferro et al. 2004).  Further, the light curve of V2 could not be phased well with this derived period. 
The present data was also not phased well with the period $\sim$ 0.49 days reported by 
 Newburn (1957), Mannino (1957), Arellano Ferro et al. (2004) and Stetson et al. (2005).
The LS power spectrum of V2 is shown in Fig. 8(a), where the prominent peak corresponds to a period of  0.335070 days, however a period of $\sim$0.49 days is also present in the data. We have made a comparison between power spectrum obtained 
using present data and that obtained from data of Arellano Ferro et al. (2004). Since the LS power spectrum is noisy, therefore, a CLEAN algorithm (Roberts et al. 1987) was  used to determine the period. Fig. 8 (b) shows the CLEANed power spectrum of present data in $V$ and $R$ bands and the $V$-band data of  Arellano Ferro et al. (2004). CLEANed power spectra were obtained by using the loop gain of 0.1 and iteration of 100. The CLEANed power spectra clearly show the period of  0.335070 from present data and $\sim$0.49 days from Arellano Ferro et al. (2004) data.  We have also applied the same approach as applied for V1 to determine the period of star V2. The LS and CLEANed power spectra of combined data are shown in Fig. 8 (c) and (d), respectively.  Both power spectra show the prominent peaks at periods  0.331181093 days and 0.493055397 days. The combined data  could not phased be well with a period of  0.493055397 days as can be seen from its light curve displayed in Fig. 9.  This indicates that both periods may be present in the data. 
Star V2 is a known RRab type variable. The period $\sim$0.33 days in the present study seems to be
too short for a RRab type variable and
this might be due to the short observational baseline.
The $\sim$0.33 days period is an alias because of the uncertainty in the
number of cycles elapsed in one day. 
A close inspection of
the light curves plotted for V2 in Fig. 5 reveals an overlap
between the descending branch and ascending branch at minimum
light.  This is very clear in the $R$ band light curve and to a lesser
extent in the $V$ band light curve.  It is an indication that the $\sim$0.33 days
period is too short.
A complicating factor in the case of V2 is that it is a Blazhko
variable.  This has been well illustrated by Arellano Ferro et
al. (2004) with their extensive data coverage.  The data in
this  present work have been obtained on only 6 nights and it
appears that observations have been obtained at different Blazhko
phases. We assume that this is the reason for the disconnect near
maximum light in the light curves plotted in Fig. 9.  In
view of this, it is not surprising that period searches with
the current data do not favour the $\sim$0.49 days period.  Also it
is not surprising that the combined data of Arellano Ferro et
al. (2004) and the current study might favour the $\sim$0.33 days period because there
are more observations in the current study compared with Arellano Ferro et al. (2004).
For further study, the period of V2 is adopted as $\sim$0.493 days.

The $V$ amplitudes of V6 and V7 are derived  to be $ \sim$1.02 and $\sim$1.09 mag, respectively. The present light curves for V7 show significant scattering in both $V$ and $R$ bands, indicating presence of the Blazhko effect (see Fig. 7).  However, Arellano Ferro et al. (2004) did not find any sign of the presence of Blazhko effect in V7.  

A large fraction of data for the stars V3 and V11 did not phase well with their derived periods (see Table 2). This suggests the presence of the Blazhko effect in these stars. 

Star V12 was classified as RRab type variable. We could detect V12 only in the $R$ band.  In the bottom panel of Fig. 7, we show the folded light curve of V12, where  the filled squares represent present data while filled circles show data from Arellano Ferro et al. (2004). Few  data points from  Arellano Ferro et al. (2004) which  are isolated from the light curve of V12, are shown by asterisk symbol in Fig. 7.  Excluding these data points, the amplitude of this star  was found to be $\sim$1.32 mag in the $R$ band, which is smaller than that derived by Arellano Ferro et al (2004).  A similar conclusion was also drawn by Stetson et al. (2005) about the light curve and amplitude of V12.

The  period  and amplitude of the star V13 derived from the present data are consistent with those of the previous studies.
Arellano Ferro et al. (2004) suggested it to be a double mode RRd star or a Blazhko type RRc. Their light curve shows a significant scatter at the maximum, whereas present light curve in the $V$ band  shows negligible scatter.  However, they also  mentioned that their observations are not sufficient to confirm whether it is an RRd star or a Blazhko type RRc type star. Stetson et al. (2005) reported that their data did not phase with the period given by Arellano Ferro et al. (2004). From the present observations, we confirm that V13 is an RRc type variable.

The present period for the star V14 was found to be  $\sim$0.26 days. The previous studies reported its period $\sim$0.52 days (Newburn 1957), $\sim$0.25 days (Arellano Ferro et al. 2004) and 0.35 days (Stetson et al. 2005).
For this star we have found slightly different period than that from Arellano Ferro et al. (2004) and Stetson et al. (2005).  The previous studies have also indicated that V14 is  pulsating with different periods.

\subsection{Newly identified}
Stars V24, V29, V32, V33, V35, V36, V42 and V47 are new variables which are located  on HB of NGC 4147.
All of them can reasonably be assumed as RR Lyrae stars but another consideration is  their amplitudes in  $V$ and
$R$ bands.  Amplitudes of pulsating variables are found as a function
of colour.
As illustrated in table 10 of the Stetson et
al. (2005), the amplitude of variability is always largest in $B$ and
smallest in $I$ band for RR Lyrae stars.
Considering the above fact along with periods and shape of the light curves, of 8 new HB stars we classify stars V29, V32, V33, V35 and V42 as RR Lyrae stars.
Their amplitudes of brightness variation in $V$-band are in the range of $\sim$0.17 to $\sim$0.62 mag, whereas their periods of variability are found to be in the range of $\sim$0.28 to $\sim$0.30 days.
The $R$ amplitudes of these variables except V31 vary from $\sim$0.12 to $\sim$0.55 mag.
Furthermore, their periods and amplitudes suggest that they could be probable RRc type variables.

In $V/(V-R)$ CMD, star V28 is located at the beginning of HB i.e very close to the red giant branch.
It has been considered as a probable member of the cluster because it is located at a radial 
distance of $\sim$0.16 arcmin from the cluster center.
Apart from its location in CMD, the period, amplitude and shape of light curve of V28 have close resemblance with those of RRc type variables. Hence, it
can be classified as RRc type variable.  
 
Star V31 is 
found to be lying almost in the center of the cluster at a radial distance of 0.047 arcmin. With $V$ magnitude of 16.547 mag, star V31 is comparable to other HB stars. Its period and $V$ amplitude are $\sim$0.282 days and 0.476 mag, respectively.
The shape of light curve, period and amplitude indicate that V31 might belong to the RRc type
population.

\begin{figure}
\includegraphics[height=8cm,angle=270]{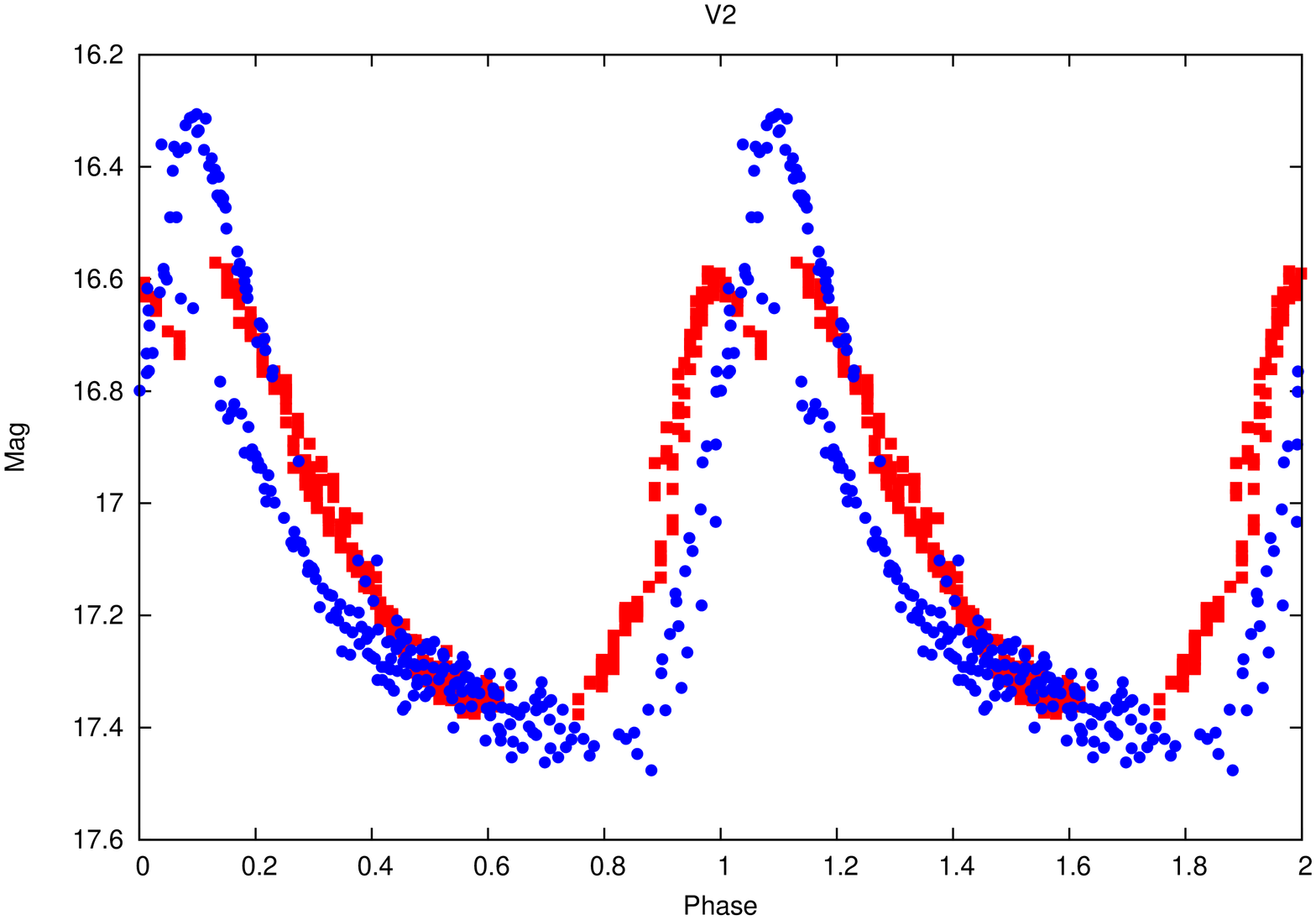}
\includegraphics[height=8cm,angle=270]{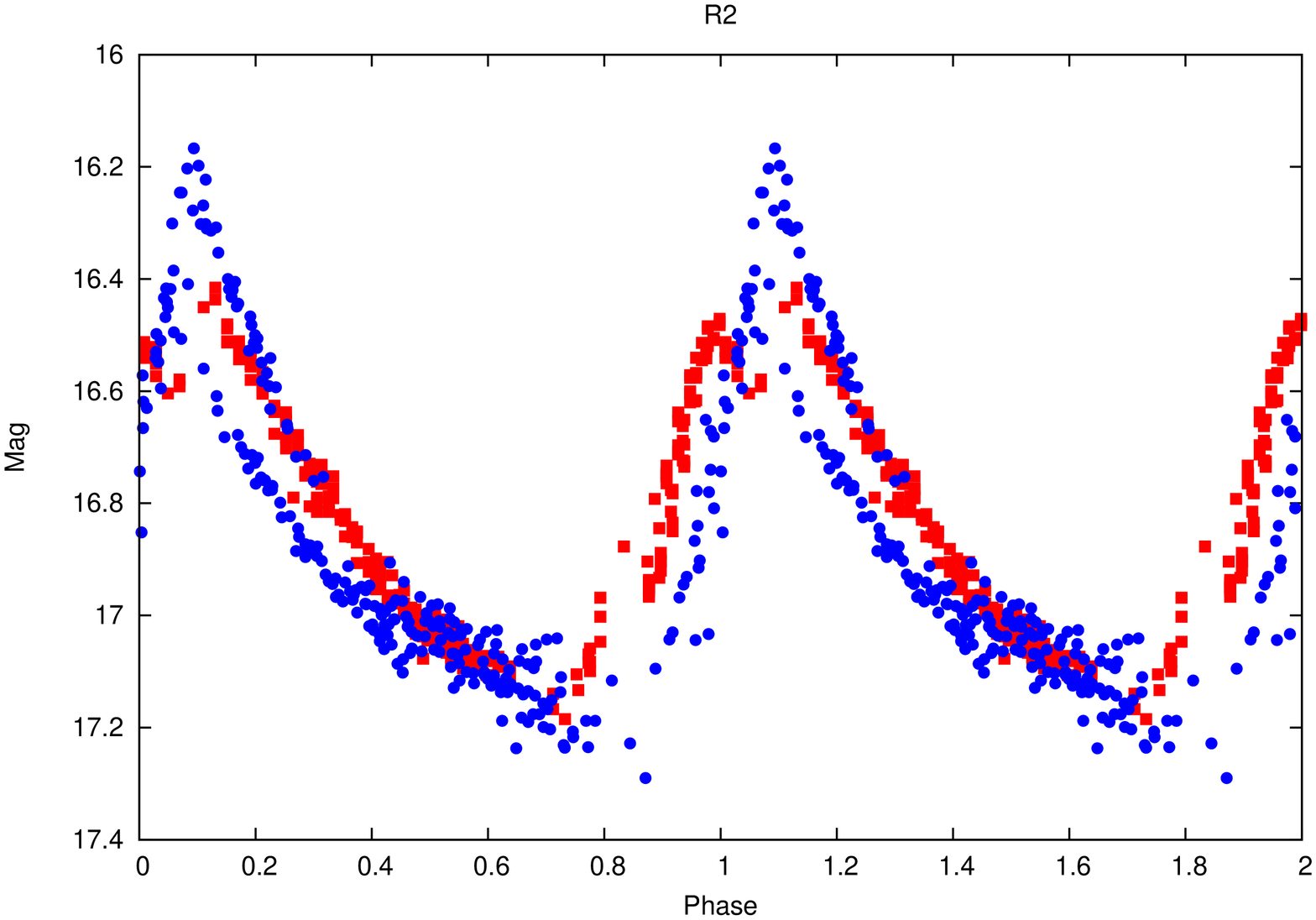}
\caption{The phased light curves of V2 in $V$ (upper panel) and $R$ (lower panel) bands folded with period $\sim$ 0.493 days. The present data are represented by filled squares while previous data (Arellano Ferro et al. 2004) are shown by filled circles.}
\end{figure}

\begin{figure}
\includegraphics[height=9cm]{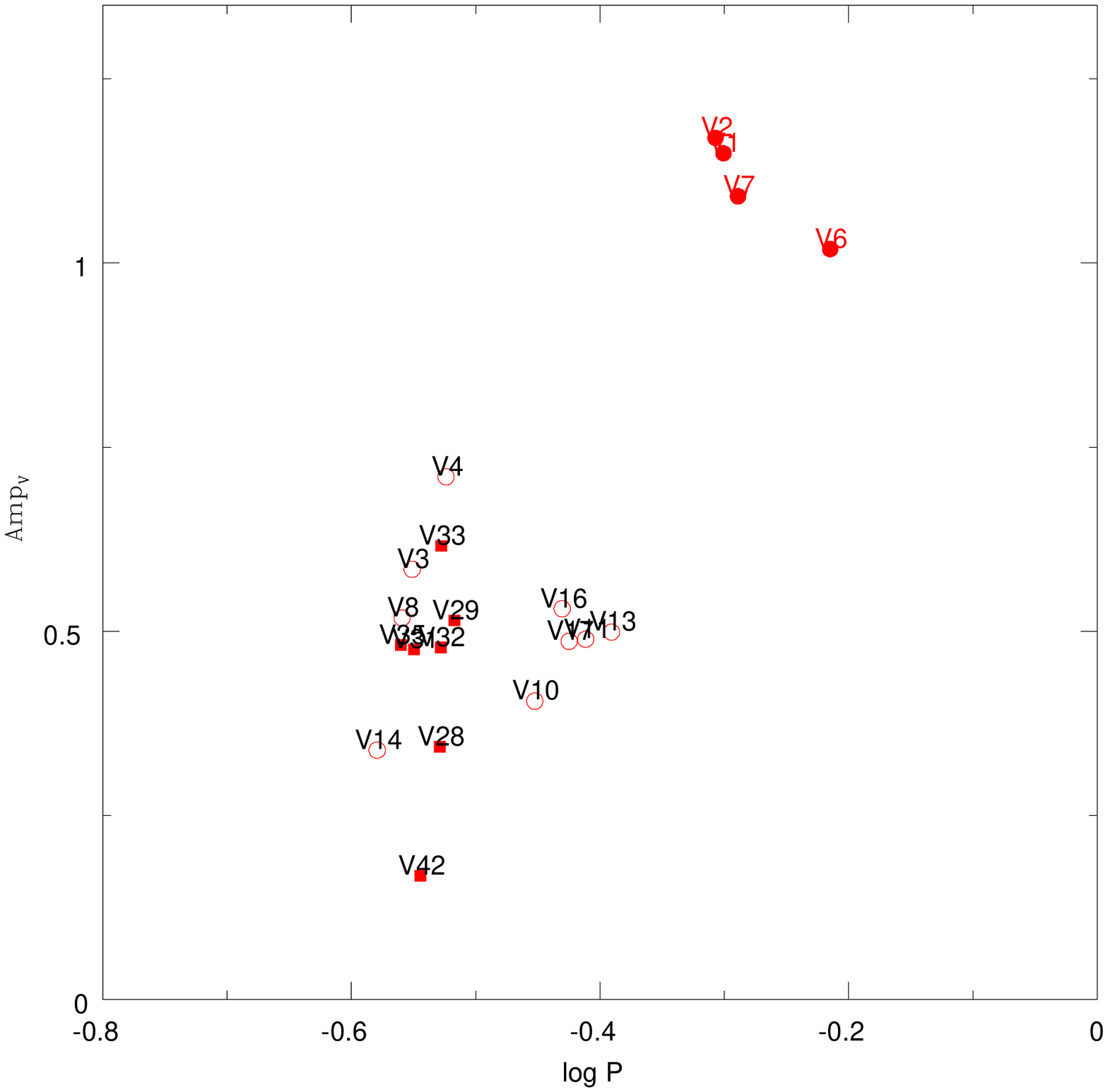}
\caption{The amplitude vs period diagram (Bailey diagram) for RRab and RRc variables. Open
and filled circles show known RRc and RRab variables, respectively. 
The newly identified RRc type variables
are shown by filled squares.} 
\end{figure}
To confirm the classification of newly identified RRc variables, we have plotted them  along with the known RR Lyrae variables except V12 in the Bailey diagram (amplitude versus period diagram) which is shown in Fig. 10. 
The filled and open circles represent known RRab and RRc pulsators, respectively, while filled squares show the newly identified RRc type variables.
As shown
in Fig. 10, the known RRab and RRc variables are
found to be distributed in two different regions, similar to previous studies in other globular clusters.
All the new RRc variables as shown in Fig. 10 are located firmly among known RRc variables, 
and this further suggests that their variability type is similar to RRc type variables. 
Here, we note that star V42
has slightly lower amplitude than that of the other RRc variables.
Hence, present study identifies seven variables as RR Lyrae stars and classify them as RRc type. 

\begin{figure}
\hbox{
\includegraphics[height=9cm,angle=270]{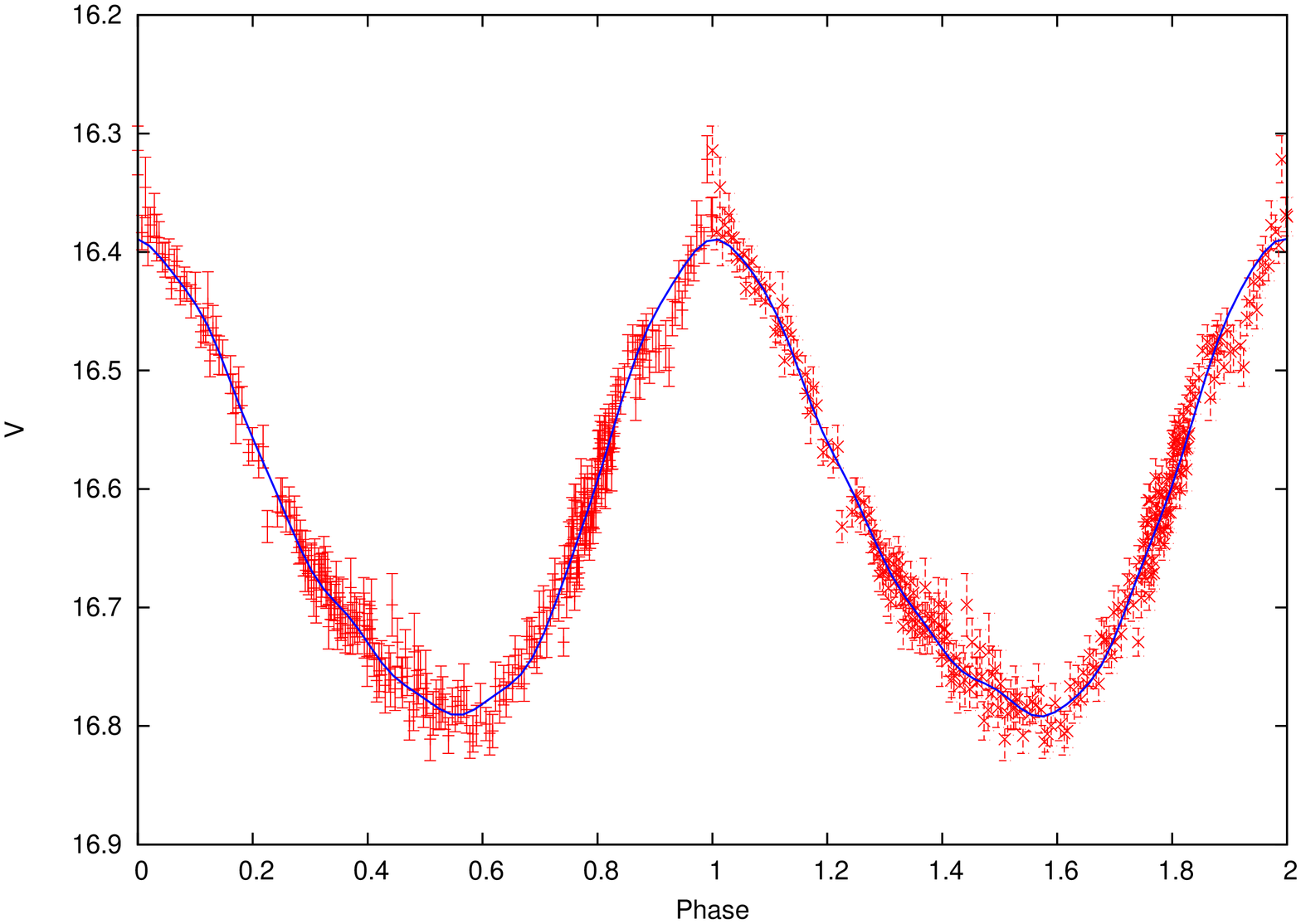}
}
\caption{The phased light curves of V13 with best fit Fourier series.}
\end{figure}

\begin{figure}
\vbox{
\includegraphics[height=9cm,angle=270]{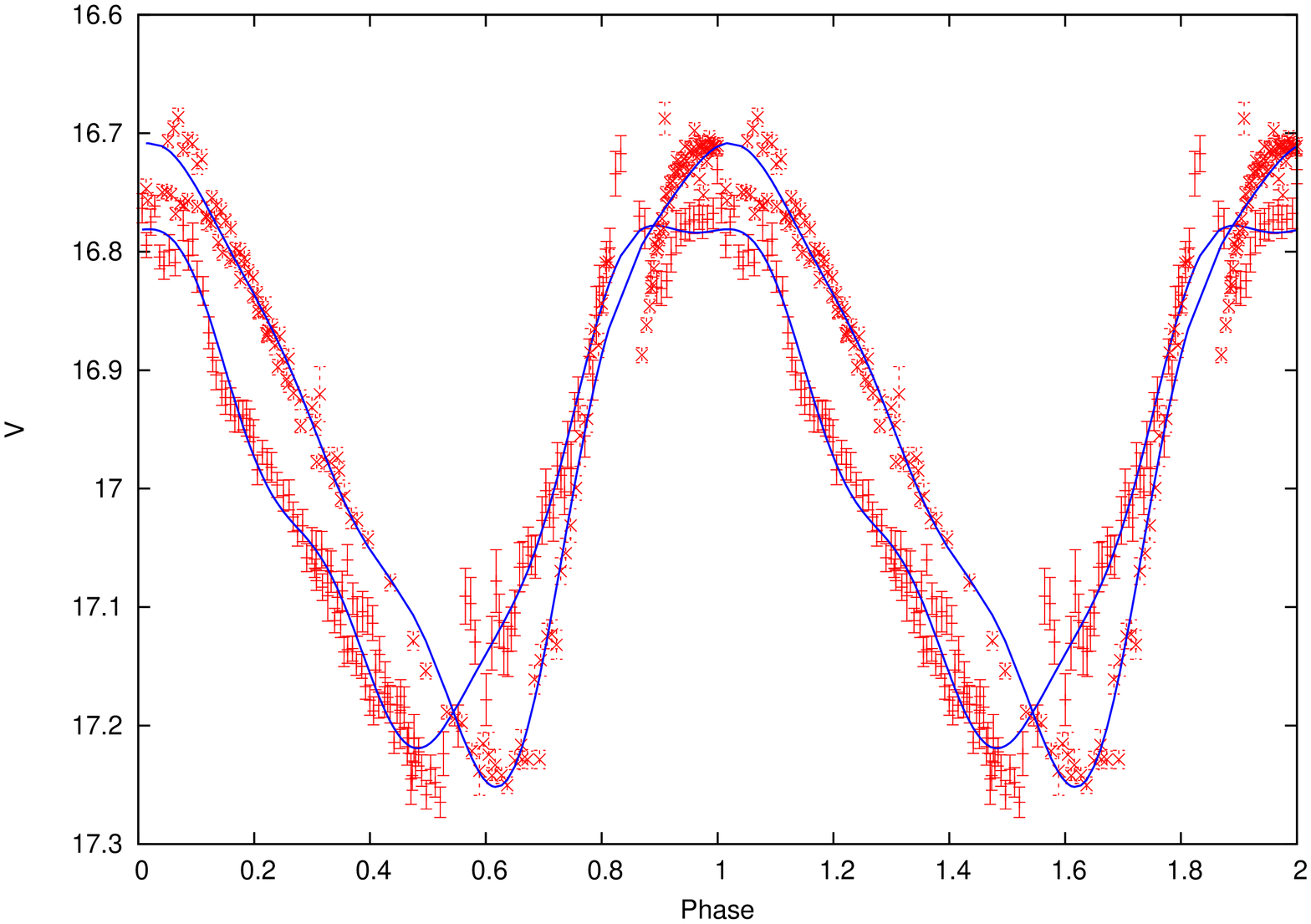}
\includegraphics[height=9cm,angle=270]{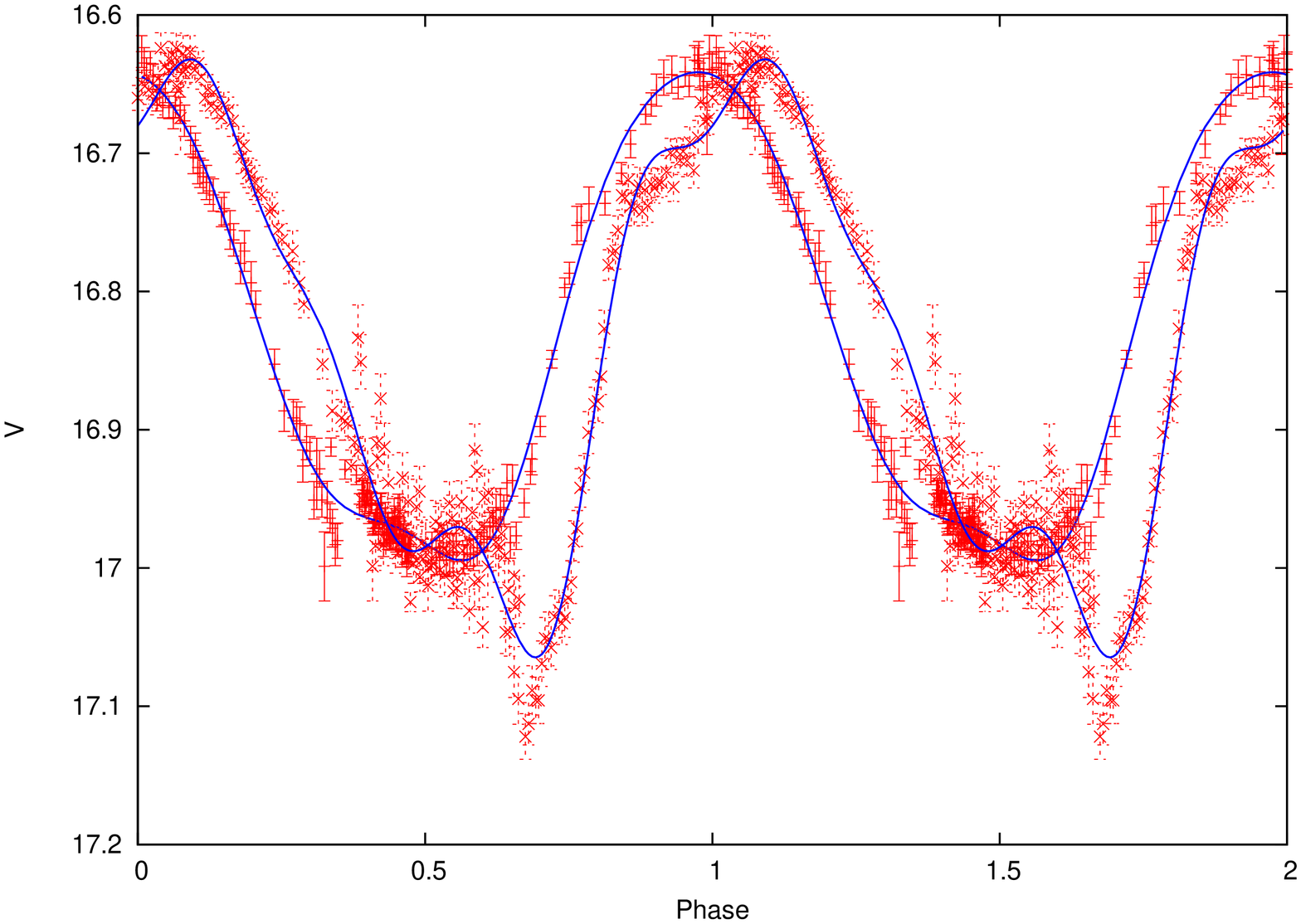}
}
\caption{The phased light curves of V3 (upper panel) and V11 (lower panel) into two parallel light curves with different nature.}
\end{figure}

\subsection{Fourier Decomposition and Physical parameters of RR Lyrae variables}
The Fourier decomposition technique consists of fitting the observed brightness variation of pulsating stars with Fourier series

$ f(\phi) = A_0 + \Sigma_{k=1}^{k=4} ~A_k ~cos(2\pi k \phi + \psi_k) $ \\
 \label{eq:fourier}

where $\phi (= (t-t_0)/P)$ is the  pulsational phase and $t_0$ is the time of maximum brightness. The best fit Fourier series for one RRc star as a sample is shown in Fig. 11, and the best-fit Fourier parameters are given in Table 4.  The structure of the light curve can be quantified in terms of combination of low-order Fourier coefficients, in particular the amplitude ratio, $R_{k1} = A_k/A_1$ ~and the phase difference $\psi_{k1} = \psi_k -k \psi_1$, where k =1,2,3,4. 
In the case of RRab variables the Fourier decomposition has done using a sine series.
A careful inspection of light curves for V3 and V11 reveals that these stars show the Blazhko effect. Therefore, we have divided  V3 and V11 data into two parallel light curves 
with different nature as shown in Fig. 12, and these variables are designated as V3(1), V3(2) and V11(1), V11(2).
The relation between the pulsation of RR Lyrae variables and their physical parameters can be obtained from hydrodynamic pulsation models with defined physical parameters, which are used to generate theoretical light curves. 
The physical parameters, 
 $[Fe/H]$, $T_{eff}$, and $M_V$ for RRab stars were calculated using 
the equations given by Jurcsik \& Kovacs (1996), Kovacs \& Jurcsik (1996) and Jurcsik (1998). These derived physical parameters for RRab variables are given in Tables 5. For RRc stars, using their phase difference parameter $\psi_{31}$, the metallicity $[Fe/H]$ is calculated from the relation 
$[Fe/H]=52.466P^2-30.075P+0.131\psi^2{_{31}}+0.982\psi_{31}-4.198\psi_{31}P+2.424 $ given by Morgan et al. (2007), 
where $P$ represents period of the variable star.  
 The calculated values of $[Fe/H]$ for RRc type stars are listed in Table 5.

\begin{table*}
\caption{ The Fourier light curve fitting parameters of individual RRab, RRc and other variables.
}
\tiny
\begin{tabular}{llllllllll}
\hline
 ID & band & A0        &  A1 & R21     &  R31     &  R41     &   $\psi$21     &  $\psi$31     &  $\psi$41                   \\        
\hline
    & &                &   &  RRab Variables &            &              &               &              &                             \\
\hline
V1  & V&      16.832$\pm$0.002  &  0.351$\pm$0.003 & 0.412$\pm$0.010   & 0.262$\pm$0.009  &  0.142$\pm$0.008   &  5.606$\pm$0.024   & 4.839$\pm$0.035  &  3.175$\pm$0.059 \\
V2  & V&      17.102$\pm$0.007  &  0.261$\pm$0.010 & 0.368$\pm$0.044  &  0.252$\pm$0.038 &   0.135$\pm$0.037  &   4.672$\pm$0.158  &  4.551$\pm$0.194  &  4.562$\pm$0.302 \\
V6  & V&      16.985$\pm$0.003  &  0.353$\pm$0.004 & 0.408$\pm$0.012  &  0.352$\pm$0.009 &   0.238$\pm$0.008  &   5.722$\pm$0.024  &  5.249$\pm$0.031  &  3.171$\pm$0.040 \\
V7  & V&      16.515$\pm$0.026  &  0.285$\pm$0.042 & 0.611$\pm$0.161  &  0.605$\pm$0.123 &   0.280$\pm$0.058  &   4.375$\pm$0.170  &  4.209$\pm$0.217  &  3.357$\pm$0.355 \\
V12 & R&      16.686$\pm$0.007  &  0.407$\pm$0.010 & 0.289$\pm$0.024  &  0.395$\pm$0.025 &   0.206$\pm$0.023  &   5.213$\pm$0.083  &  4.238$\pm$0.067  &  3.694$\pm$0.121 \\
\hline
    & &              &   &     RRc variables  &                 &                 &                  &                   & \\
\hline
 V3(1) &V &    16.979$\pm$0.002  &   0.218$\pm$0.002  &   0.097$\pm$0.011  &   0.025$\pm$0.011  &   0.091$\pm$0.011  &   1.881$\pm$0.110  &   1.509$\pm$0.412  &   3.419$\pm$0.119 \\
 V3(2) &V &    16.947$\pm$0.002  &   0.244$\pm$0.003  &   0.209$\pm$0.010  &   0.106$\pm$0.010  &   0.062$\pm$0.009  &   2.160$\pm$0.056  &   1.419$\pm$0.092  &   3.604$\pm$0.151 \\
    V4 &V &    17.017$\pm$0.003  &   0.247$\pm$0.004  &   0.291$\pm$0.015  &   0.111$\pm$0.014  &   0.068$\pm$0.013  &   3.085$\pm$0.054  &   2.911$\pm$0.135  &   1.717$\pm$0.200 \\
    V8 &V &    16.792$\pm$0.002  &   0.215$\pm$0.003  &   0.169$\pm$0.013  &   0.091$\pm$0.012  &   0.068$\pm$0.012  &   2.001$\pm$0.072  &   2.090$\pm$0.140  &   2.424$\pm$0.182 \\
   V10 &V &    16.956$\pm$0.001  &   0.165$\pm$0.002  &   0.183$\pm$0.013  &   0.029$\pm$0.012  &   0.054$\pm$0.012  &   2.367$\pm$0.076  &   2.721$\pm$0.424  &   1.988$\pm$0.231 \\
V11(1) &V &    16.827$\pm$0.001  &   0.183$\pm$0.001  &   0.118$\pm$0.007  &   0.084$\pm$0.007  &   0.012$\pm$0.007  &   3.792$\pm$0.067  &   3.174$\pm$0.090  &   3.388$\pm$0.579 \\
V11(2) &V &    16.833$\pm$0.001  &   0.188$\pm$0.002  &   0.136$\pm$0.010  &   0.118$\pm$0.009  &   0.097$\pm$0.009  &   1.308$\pm$0.082  &   3.537$\pm$0.093  &   2.515$\pm$0.104 \\
   V13 &V &    16.597$\pm$0.001  &   0.201$\pm$0.002  &   0.229$\pm$0.011  &   0.024$\pm$0.009  &   0.056$\pm$0.010  &   2.520$\pm$0.041  &   4.957$\pm$0.418  &   5.244$\pm$0.161 \\
   V14 &V &    16.665$\pm$0.002  &   0.112$\pm$0.004  &   0.118$\pm$0.031  &   0.199$\pm$0.030  &   0.079$\pm$0.027  &   2.978$\pm$0.267  &   3.243$\pm$0.167  &   3.748$\pm$0.328 \\
   V16 &V &    16.542$\pm$0.003  &   0.211$\pm$0.006  &   0.175$\pm$0.018  &   0.146$\pm$0.020  &   0.100$\pm$0.019  &   2.153$\pm$0.153  &   3.617$\pm$0.150  &   2.561$\pm$0.178 \\
   V17 &V &    16.891$\pm$0.001  &   0.237$\pm$0.002  &   0.092$\pm$0.008  &   0.091$\pm$0.006  &   0.056$\pm$0.007  &   3.576$\pm$0.073  &   2.980$\pm$0.070  &   3.133$\pm$0.108 \\
   V28 &V &    16.443$\pm$0.003  &   0.080$\pm$0.004  &   0.139$\pm$0.055  &   0.131$\pm$0.055  &   0.055$\pm$0.052  &   3.009$\pm$0.426  &   1.483$\pm$0.436  &   3.657$\pm$0.956 \\
   V29 &V &    16.971$\pm$0.004  &   0.114$\pm$0.006  &   0.389$\pm$0.056  &   0.343$\pm$0.054  &   0.135$\pm$0.054  &   2.121$\pm$0.214  &   1.650$\pm$0.227  &   2.199$\pm$0.412 \\
   V31 &V &    16.558$\pm$0.005  &   0.090$\pm$0.007  &   0.286$\pm$0.088  &   0.095$\pm$0.080  &   0.090$\pm$0.077  &   2.861$\pm$0.367  &   3.833$\pm$0.903  &   3.937$\pm$0.937 \\
   V32 &V &    16.338$\pm$0.004  &   0.133$\pm$0.006  &   0.202$\pm$0.049  &   0.120$\pm$0.044  &   0.030$\pm$0.045  &   2.481$\pm$0.254  &   2.293$\pm$0.415  &   1.942$\pm$1.415 \\
   V33 &V &    17.312$\pm$0.005  &   0.178$\pm$0.008  &   0.031$\pm$0.036  &   0.092$\pm$0.042  &   0.149$\pm$0.037  &   3.924$\pm$1.439  &   1.593$\pm$0.438  &   1.798$\pm$0.275 \\
   V35 &V &    17.073$\pm$0.005  &   0.107$\pm$0.006  &   0.407$\pm$0.064  &   0.134$\pm$0.061  &   0.073$\pm$0.059  &   2.809$\pm$0.241  &   1.730$\pm$0.492  &   2.124$\pm$0.844 \\
   V42 &V &    17.103$\pm$0.001  &   0.054$\pm$0.002  &   0.263$\pm$0.032  &   0.125$\pm$0.032  &   0.127$\pm$0.032  &   1.465$\pm$0.160  &   1.126$\pm$0.277  &   3.196$\pm$0.268 \\
\hline 
       &                       &                    &                    &           Others   &                    &                    &                    &                   \\
\hline
V20 &V  &    13.445$\pm$0.003  &   0.044$\pm$0.004  &   0.078$\pm$0.074  &   0.124$\pm$0.073  &   0.091$\pm$0.072  &   2.899$\pm$1.171  &   3.114$\pm$0.705  &   3.515$\pm$0.854 \\
V21 &V  &    18.486$\pm$0.003  &   0.076$\pm$0.004  &   0.201$\pm$0.053  &   0.259$\pm$0.054  &   0.031$\pm$0.052  &   2.164$\pm$0.306  &   2.665$\pm$0.255  &   3.918$\pm$1.620 \\
V22 &V  &    17.411$\pm$0.001  &   0.085$\pm$0.002  &   0.251$\pm$0.022  &   0.236$\pm$0.024  &   0.163$\pm$0.022  &   3.043$\pm$0.117  &   3.365$\pm$0.117  &   2.610$\pm$0.155 \\
V23 &V  &    16.713$\pm$0.001  &   0.055$\pm$0.002  &   0.260$\pm$0.025  &   0.062$\pm$0.026  &   0.130$\pm$0.026  &   3.363$\pm$0.129  &   3.578$\pm$0.417  &   3.496$\pm$0.201 \\
V24 &V  &    16.912$\pm$0.002  &   0.044$\pm$0.002  &   0.166$\pm$0.053  &   0.169$\pm$0.049  &   0.037$\pm$0.047  &   2.783$\pm$0.334  &   1.726$\pm$0.328  &   3.265$\pm$1.242 \\
V25 &V  &    17.682$\pm$0.002  &   0.036$\pm$0.002  &   0.108$\pm$0.064  &   0.036$\pm$0.059  &   0.041$\pm$0.058  &   1.708$\pm$0.567  &   2.976$\pm$1.748  &   3.360$\pm$1.522 \\
V26 &V  &    16.786$\pm$0.001  &   0.040$\pm$0.001  &   0.328$\pm$0.034  &   0.084$\pm$0.031  &   0.083$\pm$0.032  &   3.628$\pm$0.147  &   1.362$\pm$0.412  &   1.204$\pm$0.396 \\
V27 &V  &    18.313$\pm$0.004  &   0.100$\pm$0.005  &   0.177$\pm$0.051  &   0.149$\pm$0.049  &   0.179$\pm$0.050  &   3.318$\pm$0.328  &   3.062$\pm$0.376  &   2.935$\pm$0.303 \\
V30 &V  &    17.402$\pm$0.002  &   0.038$\pm$0.002  &   0.055$\pm$0.054  &   0.289$\pm$0.056  &   0.271$\pm$0.056  &   2.074$\pm$1.066  &   2.870$\pm$0.254  &   1.051$\pm$0.256 \\
V34 &V  &    19.350$\pm$0.004  &   0.149$\pm$0.006  &   0.411$\pm$0.039  &   0.176$\pm$0.036  &   0.026$\pm$0.034  &   1.017$\pm$0.137  &   4.025$\pm$0.229  &   2.378$\pm$1.376 \\
V36 &V  &    17.267$\pm$0.001  &   0.027$\pm$0.001  &   0.232$\pm$0.051  &   0.312$\pm$0.049  &   0.052$\pm$0.048  &   3.006$\pm$0.248  &   1.992$\pm$0.216  &   2.346$\pm$0.917 \\
V37 &V  &    15.695$\pm$0.001  &   0.017$\pm$0.001  &   0.460$\pm$0.053  &   0.257$\pm$0.049  &   0.157$\pm$0.046  &   1.002$\pm$0.171  &   2.906$\pm$0.228  &   1.499$\pm$0.327 \\
V38 &V  &    18.525$\pm$0.005  &   0.055$\pm$0.007  &   0.283$\pm$0.125  &   0.140$\pm$0.120  &   0.218$\pm$0.126  &   3.531$\pm$0.547  &   1.206$\pm$0.929  &   2.330$\pm$0.639 \\
V39 &V  &    18.134$\pm$0.004  &   0.082$\pm$0.006  &   0.255$\pm$0.071  &   0.045$\pm$0.068  &   0.183$\pm$0.069  &   1.844$\pm$0.331  &   3.239$\pm$1.487  &   1.711$\pm$0.417 \\
V40 &V  &    18.729$\pm$0.003  &   0.086$\pm$0.005  &   0.220$\pm$0.056  &   0.193$\pm$0.057  &   0.111$\pm$0.056  &   3.668$\pm$0.306  &   1.212$\pm$0.326  &   3.430$\pm$0.520 \\
V41 &V  &    17.508$\pm$0.001  &   0.038$\pm$0.001  &   0.150$\pm$0.042  &   0.356$\pm$0.043  &   0.092$\pm$0.039  &   4.007$\pm$0.313  &   1.181$\pm$0.179  &   2.493$\pm$0.488 \\
V43 &V  &    18.127$\pm$0.002  &   0.089$\pm$0.003  &   0.329$\pm$0.038  &   0.057$\pm$0.035  &   0.025$\pm$0.034  &   1.132$\pm$0.146  &   1.042$\pm$0.608  &   1.877$\pm$1.405 \\
V44 &V  &    18.552$\pm$0.003  &   0.060$\pm$0.004  &   0.423$\pm$0.067  &   0.282$\pm$0.061  &   0.087$\pm$0.057  &   1.596$\pm$0.218  &   2.377$\pm$0.274  &   2.844$\pm$0.697 \\
V45 &V  &    18.849$\pm$0.002  &   0.073$\pm$0.003  &   0.181$\pm$0.047  &   0.128$\pm$0.045  &   0.147$\pm$0.047  &   2.910$\pm$0.278  &   3.071$\pm$0.387  &   2.052$\pm$0.326 \\
V46 &V  &    18.424$\pm$0.004  &   0.120$\pm$0.005  &   0.436$\pm$0.051  &   0.253$\pm$0.053  &   0.159$\pm$0.049  &   1.059$\pm$0.192  &   2.739$\pm$0.234  &   1.906$\pm$0.332 \\
V47 &V  &    16.932$\pm$0.001  &   0.015$\pm$0.001  &   0.296$\pm$0.074  &   0.191$\pm$0.076  &   0.204$\pm$0.077  &   1.060$\pm$0.337  &   2.874$\pm$0.439  &   4.028$\pm$0.417 \\
\hline
\end{tabular}
\end{table*}

\begin{table}
\caption{ The physical parameters of RRab and RRc variables.
}
\scriptsize
\begin{tabular}{llll}
\hline
  ID    &   $[Fe/H]$&  $M{_V}$&   $\log$$T_{eff}$ \\
\hline
       & RRab    &       &    \\
\hline
V01    & -1.23   &  0.85 & 3.793 \\
V02    & -1.59   &  0.81 & 3.787 \\
V06    & -1.26   &  0.74 & 3.787 \\
V07    & -2.15   &  0.80 & 3.776 \\
V12    & -2.06   &  0.76 & 3.786 \\
\hline
       & RRc     &       &   \\
\hline
V3(1)  &-1.885   &-      &-  \\
V3(2)  &-1.902   &-      &-  \\
V4     &-1.567   &-      &-  \\
V8     &-1.677   &-      &-  \\
V10    &-2.044   &-      &-  \\
V11(1) &-2.079   &-      &-  \\
V11(2) &-1.994   &-      &-  \\
V13    &-1.495   &-      &-  \\
V14    &-0.883   &-      &-  \\
V16    &-1.881   &-      &-  \\
V17    &-2.081   &-      &-  \\
V28    &-1.979   &-      &-  \\
V29    & -1.999  &-      &-  \\
V31    & -0.737  &-      &-  \\
V32    & -1.794  &-      &-  \\
V33    & -1.968  &-      &-  \\
V35    & -1.491  &-      &-  \\
V42    & -1.965  &-      &-  \\
\hline
\end{tabular}
\end{table}

\subsection{Oosterhoff classification and distance determination}
NGC 4147 was classified by Castellani \& Quarta
(1987) as Oosterhoff type I in spite of having 
 a low metallicity (-1.83; Harris catalogue).
Oosterhoff (1939) divided globular clusters into two groups; Oosterhoff type I and Oosterhoff type II. The Oosterhoff type I globular clusters are found to be more metal-rich ($[Fe/H]$ $>$ $-1.7$) than
Oosterhoff type II ($[Fe/H]$ $<$ $-1.7$) (Smith 1995).
Recently, Villanova et al. (2016) presented an extensive spectroscopic study 
on the globular cluster NGC 4147 and found its metallicity $[Fe/H] = $-1.84, which is comparable to the typical metallicity of halo globular clusters.
Stetson et al. (2005) estimated the value of $[Fe/H]$ as -1.55.
The present mean value of $[Fe/H]$ for RRab variables estimated using hydrodynamic pulsation models is -1.658. For RRc stars, the present mean value of $[Fe/H]$ comes out to be -1.746, which is on the scale of Zinn \& West (1984). The iron abundance estimated for RRab stars is on the Jurcsik \& Kovacs (1996) scale, while the metallicity of NGC 4147 adopted by Harris (1996; 2010 Edition) is on the Zinn \& West (1984) scale. Therefore, to convert the present mean value of $[Fe/H]$ for RRab stars into Zinn \& West (1984) scale we have used  
a relation ($[Fe/H]_{J} = 1.431[Fe/H]_{ZW} + 0.88$) given by
Jurcsik (1995). On the scale of Zinn \& West (1984), the mean metallicity of RRab stars is, thus, translated as 
 -1.774. The present average metallicity of RRab and RRc variables
is -1.760, which is slightly smaller than that (-1.57) obtained from the cluster CMD by fitting theoretical models and found to be more close to the value provided in Harris catalogue as compared with that obtained by Stetson et al. (2005). 
The present mean $[Fe/H]$ value obtained from RRab and RRc variables seems to be consistent with that of Oosterhoff type II globular clusters.

The Oosterhoff classification of globular clusters is also discussed based on mean periods of the RRc and RRab variables, and ratio of the number of RRc to RRab variables.  For globular clusters van Agt \& Oosterhoff (1959) determined  the mean periods of the RRc and RRab variables as 0.319 and 0.549 days, respectively.
Clement et al. (2001) found the mean RRab and RRc periods for globular clusters as  0.559 days and 0.326 days, respectively, for Oosterhoff type I, and 0.659 days and 0.368 days, respectively, in the case of Oosterhoff type II. 
Based on mean periods of the RRc and RRab variables, and ratio of the number of RRc to RRab variables Stetson et al. (2005) already gave a full discussion of
the Oosterhoff classification of NGC 4147.
They found that the mean periods of the RRc
and RRab variables were characteristic of an Oosterhoff type
I cluster, but the ratio of the number of RRc to RRab variables
was characteristic of type II. 
Since present work identified 7 new RRc type variables, we have revised the
value for mean period of RRc type stars only and the ratio
of the number of RRc to RRab variables.
The calculated mean period of RRc
 is found as $\sim$0.317 days. 
The present mean period of RRc stars is in good agreement with that given by van Agt \& Oosterhoff (1959).
The number ratio of the RRc including newly identified RRc variables to the total number of RR Lyrae type is calculated as $Nc/(Nc+Nab)$= 0.76, which is larger than that (0.67) obtained by Stetson et al. (2005). 
This confirms blue HB morphology as concluded by Stetson et al. (2005).

The mean value of absolute magnitude, $M_{V}$, for RRab stars is determined to be 0.792 mag.   
From this mean $M_{V}$,  the distance modulus $(V-M_{V})$ of the cluster NGC 4147 is calculated and found to be about 16.25 mag, which gives distance to the cluster as 17.30 kpc. The mean apparent $V$ magnitude of RRab variables
is taken as 17.049 mag, which has been calculated from $V$ magnitudes given in Table 2.
The above derived distance of the cluster NGC 4147 agrees with that (17.49 kpc) obtained from the $V/(V-R)$ CMD.

\section{Other variables}
There are several studies which show that globular clusters also contain variables other than RR Lyrae stars.
These could be SX Phe, eclipsing binaries, semi-regular (SR) and other types (McNamara 1995, Kaluzny 1996, Mazur et al. 2003, Arellano Ferro et al. 2011, Kopacki et al. 2012, Kopacki 2015, Martinazzi et al. 2015). 
Recently, Kaluzny et al. (2016) and Rozyczka et al. (2017) in the case of globular clusters NGC 2301 and M22 presented numerous light curves for eclipsing binaries and SX Phe variables, and classified them based on their light curves, periods and amplitudes.
The present work contains 28 new variables, 7 of them have been classified as 
RRc variables (see section 5.2). Out of remaining 21 variables, 17 variables namely
 V21, V22, V24, V25, V27, V30, 34, 
V36, V37, V38, V39, V40, V41, V43, V45, V46 and V47 are probable members of the cluster. 
In order to classify these 17 probable member variables we have studied light curves and variability characteristics 
of several known eclipsing, SX Phe and other type variables. 
We have classified these 17 probable member variables on the basis of their light curves, period and amplitudes.
Thus, based on the light curves and other characteristics of variability, stars V21, V22, V24, V25, V27, V30, V36, V37, V39, V40, V41, V46 and V47 
could be eclipsing binaries, 
 out of which stars V24, V27, V30 and V36 are found to be located within the half light radius of the cluster.
 These stars, according to variability types
listed in the Moscow General Catalogue of Variable Stars (GCVS), are classified as Eclipsing binary systems (E), Algol (Beta Persei)-type eclipsing systems (EA) and W Ursae Majoris-type eclipsing variables (EW).
Based on the light curves, stars V24, V36, V37, V39 and V46 are classified as EA type, while stars V22, V25 and V30 are classified as E type, and variables V21, V27, V40, V41 and V47 as EW type binaries.

The shape of light curves for V34 and V43 look like RR Lyrae variables. Their variability characteristics do not match with their location 
in the $V/V-R$ CMD. Therefore, these two could be probable RR Lyrae stars which might belong to the field population, and these are termed as 
suspected field stars. 
In section 3.2, V34 and V43 were considered to be probable members of the cluster on the basis of their geometric probabilities, 
where we did not consider variability type.  
We could not classify star V45 on the basis of its light curve, period and amplitude. 

The probable member variable, V38, is found to be located at 0.846 arcmin away from the cluster center, and  in $V/(V-R)$ CMD it lies
in the location of blue stragglers. The location in the $V/(V-R)$ CMD, period and amplitude of V38 suggest 
that it could be a probable
candidate of SX Phe type variables. 

Stars V20, V23, V26 and V44 are identified as field stars based on their geometric probabilities. 
Among these four stars, V20 and V44 seem to be EA type variable based on
their light curves.  

We could not detect any SR type variable in the present cluster, and this might be due to the short span of the observations.
In the CMD of globular clusters, SR type variables are located near the top of the red giant branch, brighter than the HB. According to the GCVS, they pulsate with period  from tens of days to more.  
A good discussion of the SR variables in M13 was recently
presented by Osborn et al. (2017) in which they have shown CMD positions, periods and
classifications of SR variables. 

\begin{figure*}
\includegraphics[height=19cm]{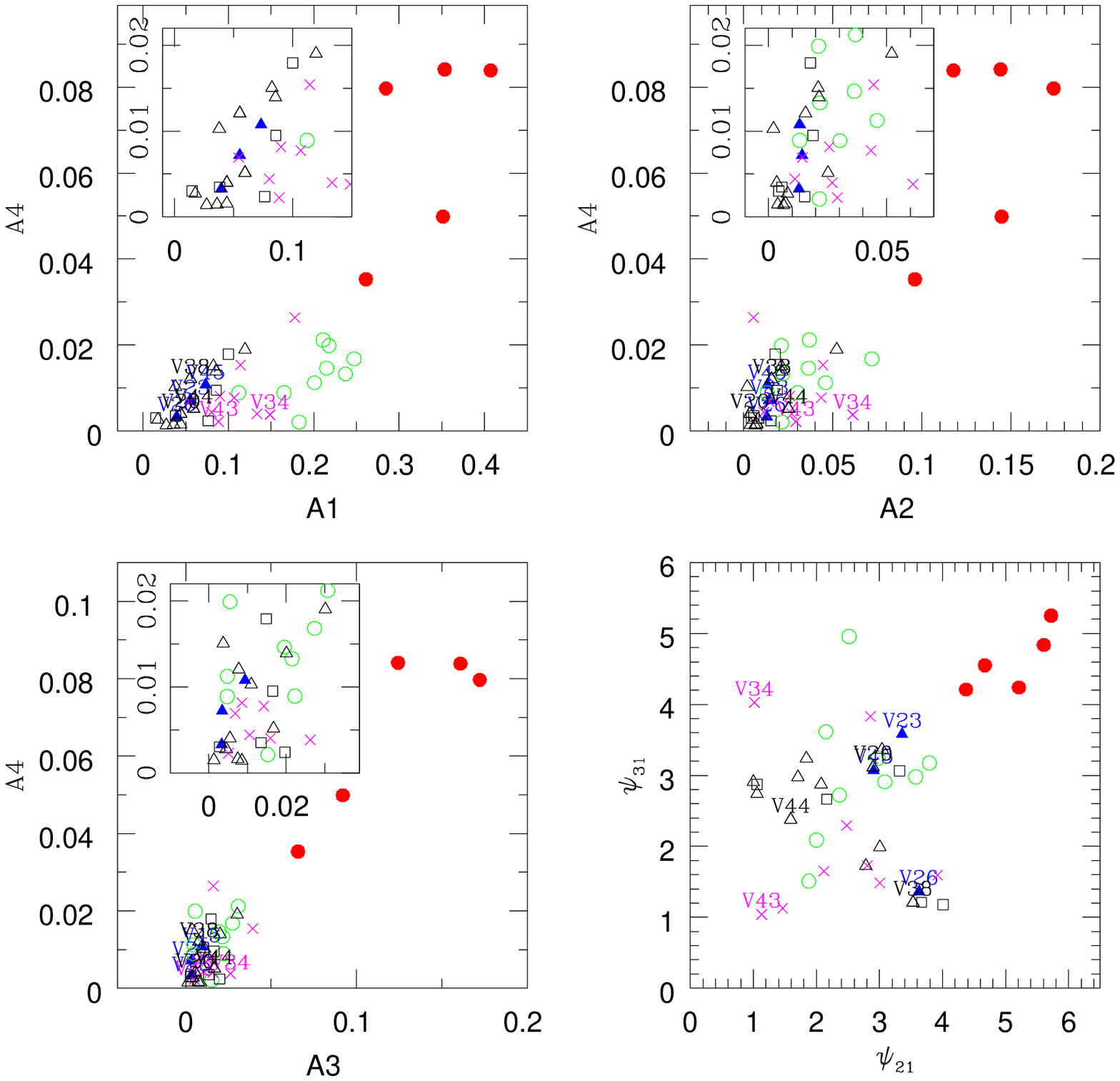}
\caption{Relative Fourier parameters for all the variables identified in the present work. The different symbols used are as filled circle (known RRab), open circle (known RRc), cross (new RRc type including two suspected field stars V34 and V43),
open triangle (EA/E including SX Phe and two field stars V20 and V44), square (EW), and filled triangle (V23, V26 and V45).  The crowded region at left corner of the amplitude ratio plots is zoomed in the inset of these plots.}
\end{figure*}

\subsection{Classification based on Fourier parameters}
Automated classification of variable stars is necessary
because of having large data sets.
Recently, Graham et al.  (2017) discussed the challenges in the automated
classification of variable stars.
A useful tool for automated
classification is
Fourier decomposition of light curves.
Nowdays, the Fourier decomposition
technique is used to distinguish between eclipsing binaries and RR
Lyrae variables.  An illustration of this technique is a paper by
Masci et al. (2014) which is on  automated classification
of periodic variable stars. Their paper deals with RR Lyrae,
EA and EW type variables and they
use the Fourier phase difference, $\phi_{21}$ and $\phi_{31}$, as well
as the amplitudes A2 and A4. 

In view of above, to see the distribution and verify present classification of variables 
we
have plotted all the present identified variables in various combination of their Fourier 
parameters in Fig. 13.
The Fourier parameters of all the present eclipsing binaries, field stars, and unclassified star V45 including one SX Phe have been 
estimated in the similar manner as discussed in section 5.3 and listed in Table 4.
The A1/A4 diagram of Fig. 13 shows that RR Lyraes and eclipsing binaries are
populated in the different regions.
In Fig. 13, the A1/A4 diagram shows more clear separation of RR Lyraes and eclipsing binaries in comparison to the A2/A4 and A3/A4 diagrams. The A2/A4 diagram segregates different type of variable stars more than the A3/A4 diagram but not to that extent as the A1/A4 diagram.
In present study, it seems that EW type variables are located with EA type in the A1/A4, A2/A4 and A3/A4 diagrams, but
in figure 3 of Masci et al. (2014), within given range 0.05$<$A2$<$0.02 and 0.05$<$A4$<$0.01, EW type variables are found to be distributed among EA type variables.
The distribution of present EA, EW and RRc type variables seems to be consistent with that given by Masci et al. (2014).
It seems that the present classification of eclipsing binaries based on their light curves is reasonable. 
The $\psi_{31}/\psi_{41}$ diagram shows that new RRc member variables are located among previously known RRc type variables, while
from this diagram it is very difficult to discuss the nature of present eclipsing binaries.

\section {Summary}
The $VR$ CCD photometry and detection of 42 periodic variables in the region of NGC 4147 are presented using 
data taken from 3.6-m Devasthal Optical telescope (DOT), Aryabhatta Research Institute of 
Observational Sciences (ARIES), India. For the majority of the known variables,
we present better light curves. 
Twenty eight new variables have been detected in the present work, and most of them belong to the HB and red giant branch.
Ten new variable stars are distributed within the half light radius $\lesssim$ 0.48 arcmin of the cluster, of which two stars are located within the core radius $\lesssim$ 0.09 arcmin.
The membership of variable candidates has been established on the basis of $V/(V-R)$ CMD, proper motions and geometric probabilities, and majority
of them are found to be members of the cluster. 
Seven newly identified probable members of the cluster could be RRc type,  while we have classified 8 new members as AE/E, 5 as EW type eclipsing
binaries and 1 as SX Phe type variable.
The iron abundance estimated from RRab
 and RRc type variables is indicative of NGC 4147 being an Oosterhoff type II cluster. 
The distance modulus of the cluster derived from light curves of RRab variables using Fourier decomposition technique 
is $\sim$16.25 mag which agrees with that(16.40 mag) obtained from the cluster $V/(V-R)$ CMD by fitting theoretical models. 

\section{Acknowledgment}
Authors are thankful to the anonymous referee for his/her critical suggestions/comments.
We are thankful to the governing council chairmen, Prof. K Kasturirangan, Prof. S. K. Joshi, Prof. G. Swarup and all the PMB members
for their guidance to successfully install 3.6 m DOT. 
Authors are highly grateful to the former director of ARIES, Prof. Ram Sagar for his significant contribution towards 3.6 m DOT. 
We thank our colleagues, Dr. Maheswar Gopinathan, Dr. Amitesh Omar, Dr. Alok C. Gupta and Dr. Santosh Joshi, for his contribution in 3.6 m DOT project.
Authors acknowledge all the technical, administrative and supporting staff of ARIES
for their wholehearted support in the realization of the 3.6 m DOT project.

\section*{References}
\noindent
Arellano Ferro A., Figuera Jaimes R., Giridhar S., Bramich D. M., Hern{\'a}ndez Santisteban J. V., Kuppuswamy K., 2011, MNRAS, 416, 2265\\
Arellano Ferro A., Ar{\'e}valo M. J., L{\'a}zaro C., Rey M., Bramich D. M. and Giridhar S., 2004, Rev. Mex. A\&A, 40, 209\\
Baade W., 1930, Astron Nachr., 239, 353\\
Baker JoDee M., Layden Andrew C., Welch Douglas L., Webb Tracy M. A., 2007, AJ, 133, 139\\
Cabrera-Cano J., Alfaro, E. J., 1990, A\&A, 235, 94 \\
C{\'a}ceres C., Catelan M.,  2008, ApJS, 179, 242\\
Castellani V., Quarta M. L., 1987, A\&AS, 71 103001\\
Clement, 2017, In Wide-Field Variability Surveys: A 21st Century Perspective, published in EPJ Web of Conferences, Volume 152, 01021\\
Clement C., 1997, A Third Catalogue of Variable Stars in Globular Clusters;\\
http://www.astro.utoronto.ca/cclement/cat/clusters.html\\
Clement Christine M., Muzzin Adam, Dufton Quentin, Ponnampalam Thivya, Wang John, Burford Jay, Richardson Alan, Rosebery Tara, Rowe Jason, Hogg Helen Sawyer,  2001, AJ, 122, 2587\\
Clementini G., Gratton Raffaele G., Bragaglia A., Ripepi V., Martinez F., Aldo F., Held Enrico V., Carretta E., 2005, ApJ, 630, 145\\
Corwin T. M., Sumerel Andrew N., Pritzl Barton J., Smith Horace A., Catelan M., Sweigart Allen V., Stetson Peter B., 2006, AJ, 132, 1014\\
Davis H., 1917, PASP, 29, 260\\
Gaia Collaboration et al.  , 2018 , A\&A , 616 , 13 https://doi.org/10.1051/0004-6361/201832900\\
Graham M., Drake A., Djorgovski S. G., Mahabal A., Donalek Ciro, 2017, In Wide-Field Variability Surveys: A 21st Century Perspective, published in EPJ Web of Conferences, Volume 152,\\
Girardi L., Bertelli G., Bressan A., Chiosi C., Groenewegen M. A. T., Marigo P., Salasnich B., Weiss A., 2002, A\&A, 391, 195\\
Harris W. E., 1996, AJ, 112, 1487\\
Jurcsik J., 1998, A\&A, 333, 571\\
Jurcsik J., Kovacs G.,  1996, A\&A, 312, 111\\
Jurcsik J. 1995, AcA, 45, 653\\
Kovacs G., Jurcsik J., 1996, ApJ, 466, L17\\
Kopacki G.,  2015, AcA, 65, 81\\
Kopacki G., Pigulski A.,  2012, arXiv1211.5465\\
Kopacki G., Drobek D., Ko{\l}aczkowski Z., Po{\l}ubek G., 2008, AcA, 58, 373\\
Kaluzny J., Rozyczka M., Thompson I. B., Narloch W., Mazur B., Pych W., Schwarzenberg-Czerny A.,  2016, AcA, 66, 31\\
Kaluzny J., 1996, A\&AS, 120, 83\\
Layden Andrew C., Bowes Benjamin T., Welch Douglas L., Webb Tracy M. A., 2003, AJ, 126, 255\\
Lomb N. R., 1976, ApSS, 39, 447\\
Masci Frank J., Hoffman Douglas I., Grillmair Carl J., Cutri Roc M,  2014, AJ, 148, 21\\
Mazur B., Krzemi{\'n}ski W., Thompson Ian B., 2003, MNRAS, 340, 1205\\
Martinazzi E., Kepler S. O., Costa J. E. S., Pieres A., Bonatto C., Bica E., Fraga L., 2015, MNRAS, 447, 2235\\
McNamara D. H., 1995, AJ, 109, 1751\\
Mannino G., 1957, Mem. Soc. Astron. Italiana, 28, 285\\
Morgan Siobahn M., Wahl Jennifer N., Wieckhorst Rachel M., 2007, MNRAS, 374, 1421\\
Newburn R. L., 1957, AJ, 62, 197\\
Oosterhoff P. Th., 1939, Observatory, 62, 104\\
Osborn W., Layden A., Kopacki G., Smith H., Anderson M., Kelly A., McBride K., Pritzl B., 2017, AcA, 67, 131\\
Pandey S. B., Yadav R. K. S., Nanjappa Nandish, Yadav Shobhit, Krishna Reddey B., Sahu Sanjeet, Srinivasan R.,  2017, arXiv171105422P\\
Pritzl Barton J., Smith Horace A., Catelan M., Sweigart Allen V., 2002, AJ, 124, 949\\
Roberts D. H., Lehar J., Dreher J. W., 1987, AJ, 93, 968\\
Rozyczka M., Thompson I. B., Pych W., Narloch W., Poleski R., Schwarzenberg-Czerny A.,  2017, AcA, 67, 203\\
Sandage A. R., Walker M. F., 1955, AJ, 60, 230\\
Scargle J. D., 1982, ApJ, 263, 835\\
Smith H. A., 1995, Sci, 270, 123\\
Stetson Peter B., Catelan M., Smith Horace A., 2005, PASP, 117, 1325\\
Stetson P.~B, 1992, J. R. Astron. Soc. Can., 86, 71\\
Stetson P.~B., 1987, PASP, 99, 191\\
Valcarce A. A. R., Catelan M., De Medeiros J. R., 2013, 553, A\&A, 62\\
Valcarce A. A. R., Catelan M., Sweigart A. V., 2012, A\&A, 547, 5\\
van Agt S.,  Oosterhoff P. Th., 1959, Ann. Sternw. Leiden, 21, 253\\
Villanova S., Monaco L., Moni Bidin C., Assmann P., 2016, MNRAS, 460, 2351\\
Vasilevskis S., Klemola A., Preston G., 1958, AJ, 63, 387\\
Wang J. J., Chen L., Wu Z. Y., Gupta A. C., Geffert M., 2000, A\&AS, 142, 373\\
Zinn R., West M. J., 1984, ApJS, 55, 45 (ZW84)\\

\end{document}